\documentclass[11pt]{article} 
\usepackage{amsmath}
\usepackage{amssymb}
\usepackage{latexsym}
\usepackage{cite}
\usepackage{epsfig,psfrag}
\begin{document} 
\title{Modular Localization of Massive Particles with ``Any'' Spin  in d=2+1}
\author{ Jens Mund\thanks{
Institut f\"ur Theoretische Physik, Universit\"at G\"ottingen, 
Bunsenstr.\ 9, 37\,073 G\"ottingen, Germany. E-mail: 
{\tt mund@theorie.physik.uni-goettingen.de} }} 
\date{January 2003}
\maketitle 

{\em Dedicated to Rudolf Haag on the occasion of his
  80${}^{\text{th}}$ birthday.}

\begin{abstract}
We discuss a concept of particle localization which is motivated from
quantum field theory, and has been proposed by Brunetti, Guido and
Longo and  by Schroer. 
It endows the single particle Hilbert space with a family of real 
subspaces indexed by the space-time regions, with certain specific properties
reflecting the principles of locality and covariance. 
We show by construction 
that such a localization structure exists also in the case of 
massive anyons in $d=2+1$, {\it i.e.\ }for particles 
with positive mass and with arbitrary spin $s\in\mathbb{R}$. 
The construction is completely intrinsic to the corresponding ray 
representation of the (proper orthochronous) Poincar\'e group. 
Our result is of particular interest 
since there are no free fields for anyons, which would fix a localization 
structure in a straightforward way.  
We present explicit formulas for the real 
subspaces, expected to turn out useful for the construction of a
quantum field theory for anyons. 
In accord with well-known results, only localization in string-like, 
instead of point-like or bounded, regions is achieved. 
We also prove a single-particle PCT theorem, exhibiting a PCT operator which 
acts geometrically correctly on the family of real subspaces.  
\end{abstract}
\newcommand{\Bb}{\mathbb{R}}
\newcommand{\Bi}{\mathbb{Z}}
\newcommand{\Bn}{\mathbb{N}}
\newcommand{\Bc}{\mathbb{C}}
%\renewcommand{\theequation}{\thesection.\arabic{equation}}
%
%%      Theoreme    (fuer Amsart)  %%%%%%%%
\newenvironment{Proof}%
{\par \medskip \noindent {\em Proof.}}{\hfill $\square$ %\par 
\medskip}
\newenvironment{Proofof}%
{\par \medskip \noindent {\em Proof\, }}{\hfill $\square$ %\par 
\medskip}
\newtheorem{Thm}{Theorem}[section]
\newtheorem{Prop}[Thm]{Proposition}
\newtheorem{Lem}[Thm]{Lemma}
\newtheorem{Cor}[Thm]{Corollary}
\newtheorem{Def}[Thm]{Definition}
\newtheorem{DefLem}[Thm]{Definition and Lemma}
\newenvironment{Remark}{\par\noindent {\em Remark. } \rm}{\par\medskip}
\newcommand{\ddt}{\frac{{\rm d}}{{\rm d}t}_{|_{t=0}}} 
\newcommand{\calH}{{\mathcal H}}
\newcommand{\BiWi}{Bi\-so\-gna\-no-Wich\-mann pro\-per\-ty } 
%% Algebra + Analysis 
\newcommand{\Emass}{E^{m,s}}     % Projektor auf Einteilchenraum. 
\newcommand{\field}{\varphi}     % Anyon field 
\newcommand{\lsp}{\left(\,}
\newcommand{\rsp}{\,\right)}
\newcommand{\U}{{U}}      % Rep of Poinc. 
\newcommand{\Se}{S}      % S(W_1)
\newcommand{\Tom}{\hat{S}}    % Tomita Operator 
\newcommand{\PCT}{U(\jtild)}  % PCT Operator 
\newcommand{\K}{K}          % ER von S
\newcommand{\KeWe}{\Ke(\We)}  % K(W_1)
\newcommand{\Ke}{K}           % ER von Se (auf Einteilchenraum) 
\newcommand{\CC}{C}    % charge conjugation operator  
\newcommand{\lin}{l}          % ``lineare''  Abb auf H_m.  
\newcommand{\calA}{\mathcal{A}}          % ``lineare''  Abb auf H_m.  
\newcommand{\calF}{\mathcal{F}}          % ``lineare''  Abb auf H_m.  
\newcommand{\Z}{Z}    % Twist number = exp{i pi s}
\newcommand{\strip}{G}    % Strip R + i pi 
\newcommand{\Potild}{\tilde{P}_+^{\uparrow}}
\newcommand{\Potildj}{\tilde{P}_+}
\newcommand{\Po}{P_+^{\uparrow}}
\newcommand{\Poj}{P_+}
\newcommand{\Lortild}{\tilde L_+^\uparrow}
\newcommand{\Lor}{L_+^{\uparrow}}
\newcommand{\half}{{\frac{1}{2}}} 
\newcommand{\gtild}{{\tilde{g}}}
\newcommand{\act}{\!\cdot\!}
\newcommand{\clo}{ {\mbox{\bf --}} }
\newcommand{\jtild}{{\tilde{j}}} 
\newcommand{\Wtild}{{\tilde{W}}}
\newcommand{\dom}{{\rm dom }} 
\newcommand{\Ctild}{\tilde{C}}
\newcommand{\Hyp}{H_m}
\newcommand{\pihalf}{{\frac{\pi}{2}}} 
\renewcommand{\d}{{\rm d}}
\newcommand{\supp}{{\rm supp}}
\newcommand{\eps}{\varepsilon}
\newcommand{\bfp}{{\boldsymbol{p}}}
\newcommand{\unity}{{\setlength{\unitlength}{1em}
                     \begin{picture}(0.75,1)
                     \put(0,0){$1$}
                     \put(0.34,0){\line(0,1){0.65}}
                     \end{picture}
                   }}
\newcommand{\einsi}{\frac{1}{i}}
\newcommand{\utilde}[1]{\mathop{#1}\limits_{\widetilde{\phantom{\textstyle
 #1}}}}
\newcommand{\unit}{{\mbox{\texttt 1}}} 
%%%%%%%%%%%%%%%%%%%%%%%%%%%%%%%%%%%%%%%%%%%%%%%%%%%%%%%%%%%%%%%%%%%%%%%%%%
\newcommand{\coc}{c}            % Cocycle  
%%%%%%%%%%%%%%%%%%%%%%%%%%%%%%%%%%%%%%%%%%%%%%%%%%%%%%%%%%%%%%%%%%%%%%%%%%%%
%% Geometrie 
\newcommand{\We}{\Wtild_1}             % ~W_1 
\newcommand{\spc}{{C}}             % spacelike cone 
\newcommand{\Spc}{\mathcal{C}}         % Set of spacelike cones 
\newcommand{\spcpath}{\tilde{\spc}}    % path of spacelike cones 
\newcommand{\Spcpaths}{\tilde{\Spc}}
                              % Set of paths of spacelike cones 
\newcommand{\ccc}{{C}}  % convex, causally complete, string-like region 
\newcommand{\Ccc}{\mathcal{C}}         % Set of ccc regions 
\newcommand{\cccpath}{\tilde{\ccc}}    % path of ccc
\newcommand{\Cccpaths}{\tilde{\Ccc}}   % Set of paths of ccc regions
\newcommand{\spd}{e}                   %spacelike direction 
\newcommand{\Spd}{H}              %set of spacelike directions   
\newcommand{\spdpath}{\tilde{e}}       % Weg von raumartigen Richtungen 
\newcommand{\refspd}{{e_0}}            % raumartiger Referenzvector 
\newcommand{\refspdpath}{\spdpath_0}  % raumartiger Referenzvector (Weg)  
%% Gruppenelemente: 
\newcommand{\lamtild}{{\tilde{\lambda}}} 
\newcommand{\lortild}{\lamtild}  %aus Ueberlagerungsgruppe von Lor. 
\renewcommand{\lor}{\lambda}     %aus Lorentzgruppe.
\newcommand{\potild}{\tilde{g}}  %aus Ueberlagerungsgruppe von Poinc. 
\newcommand{\po}{g}              %aus Poinc. 
\renewcommand{\lor}{\lambda}     %aus Lorentzgruppe.
\newcommand{\rot}[1]{\tilde{r}(#1)}  %Rotation in Ueberlagerung 
\newcommand{\Rot}[1]{r(#1)}          %Rotation in Lorentzgruppe 
\newcommand{\boox}[1]{\tilde{\lambda}_1(#1)}   % 1-Boost in Ueberlagerung 
\newcommand{\Boox}[1]{\lambda_1(#1)}           % 1-Boost in Lorentzgruppe 
\newcommand{\boo}[2]{\tilde{\lambda}_{#1}(#2)} % #1-Boost in Ueberlagerung 
\newcommand{\Boo}[2]{\lambda_{#1}(#2)}         % #1-Boost in Lorentzgruppe 
\newcounter{Lis}          
\newenvironment{liste}{
\begin{list}{{\rm (\theLis)}}{
\setlength{\itemsep}{0ex}  
\setlength{\parsep}{\parskip}  
 \usecounter{Lis}} 
}{\end{list}} 
\section{Introduction} 
Following E.~Wigner~\cite{Wig}, the state space of an elementary
relativistic 
particle  corresponds to an irreducible ray representation of the  
Poincar\'e group\footnote{We shall not be concerned with the 
concept of infra-particles~\cite{Bu87,S63}.}. 
In three as well as in four dimensional space-time, the physically
relevant representations -- and hence the conceivable particle types -- 
are classified by the mass $m$ and the spin $s$  which
labels  a representation of the covering of the rotation
subgroup (if $m>0$). 
 In three 
dimensional space-time the latter is isomorphic to the group of reals, 
hence the spin  may take {\em any} real value  ---  in contrast to 
the four-dimensional situation where it is quantized, $s\in\half\Bn_0$. 
Thus, in three dimensional space-time there are {more } particle
types; the exotic ones with non-half-integer spin are called {\em anyons}. 

By modular localization of particles we mean a concept which has 
been advocated in recent years by Brunetti, Guido and Longo~\cite{BGL} 
and by B.\ Schroer \cite{S97a,FasS02}: 
Suppose there is a quantum field for the particle type at hand, and
consider the single particle states which are, together with a 
polarization cloud, created from the vacuum in a given space-time
region. Thus the single particle space gets equipped with a family of subspaces
indexed by the space-time regions, with certain specific properties 
reflecting the localization properties of the underlying quantum
field, {\it cf.\ }Definition~\ref{DefLS} below. 
This will be a sufficient motivation for us to call a family of 
subspaces of the single particle space with such properties a 
{\em localization structure} for the particle type at hand. 

The question arises whether such a structure can be 
constructed for any given particle type $(m,s)$ intrinsically within
the single particle theory -- that is to say, without referring to a
quantum field, but using as input only the 
corresponding ray representation of the Poincar\'e group. 
This has been achieved for spin zero and positive
mass by P.~Ramacher~\cite{Pablo}, and for all positive energy 
representations of the Poincar\'e group by Brunetti, Guido and 
Longo~\cite{BGL}. The latter analysis includes reducible representations, 
but restricts to proper (not ray) representations, {\it i.e.\ }only 
the case of bosons and not the case of fermions or anyons is
covered. 

In the present article, this construction is performed for the case of
massive anyons in $d=3$. 
The purpose of this construction is twofold: Firstly, it shows that a
localization structure indeed exists for all $m>0, s\in\Bb$. This is of
particular interest because there are no free relativistic fields for
anyons\footnote{creating finitely many copies of the irreducible
  representation from the vacuum. 
D.R.Grigore has constructed free fields in $d=2+1$ for any spin 
\cite{Grig}, but in contradiction to the generalized spin 
statistics connection holding in algebraic quantum field 
theory~\cite{F89,FM1,BuEp} they have bosonic statistics. Presumably, this 
is due to the fields having infinitely many components.}~\cite{M}, 
which would of course allow for a straightforward 
construction of the localization structure. Even worse, none of the
hitherto proposed models of relativistic quantum fields for anyons in 
(continuous) three dimensional 
space-time~\cite{Semenoff,Swanson,Jack,Ban,Ply,Ito01,Mintchev91} has
been worked out to the extent that the localization structure could be
readily constructed from them. 
Secondly, our analysis is intended to be a step in the
construction of a model which resembles as closely as possible a free
field for anyons, in the sense of a ``second quantization functor'' 
from the single particle theories to field algebras. 
To this end it is gratifying that we have found explicit formulas for 
the real subspaces of localized states. 

The article is organized as follows. In Section~\ref{secDefLS}, we
make precise our definition of a localization structure for anyons, 
{\it cf.\ }Definition~\ref{DefLS}. 
In Section~\ref{secCon}, we construct a localization structure for any 
given particle type $m>0,s\in\Bb$, intrinsically within the 
corresponding Wigner space. The result is summarized in the main 
Theorem~\ref{LS}, which also contains a PCT theorem. 
All relevant properties can be shown,  via modular
theory along the lines of ~\cite{BGL},  
without reference to the specific irreducible representation $(m,s)$ 
--- except for  the the so-called standard property, which
guarantees that the constructed structure is non-trivial. 
This is the content of the last Section~\ref{secSta}, where we 
explicitly exhibit sufficiently many ``localized states'' 
(Proposition~\ref{SFT}). These are represented as families of 
functions which transform covariantly under the Poincar\'e group
(Corollary~\ref{SFTg}).  In Section~\ref{UniSpiSta}, we finally prove 
that the \BiWi essentially fixes the localization structure and also
implies a single-particle version of the spin-statistics connection. 
\section{Definition of a Localization Structure for Anyons} \label{secDefLS} 
Let $\calH$ be a Hilbert space describing anyons of the type
$(m,s)$. 
We define a localization structure as a family of subspaces of $\calH$ 
with certain specific properties reflecting the localization
properties of a hypothetical underlying quantum field. 

Let us first describe the index set for this family. 
Each subspace is labelled by a space-time region belonging to a specific 
class $\Ccc$, together  with 
some additional information, which is needed to endow the index set
with a partial order relation and with a non-trivial action of 
the $2\pi$-rotation. 
In accord with the well-known result~\cite{FM1,F89} that anyons cannot
be localized in point-like, but only in string-like regions, each  
localization region $\ccc\in\Ccc$ must extend to infinity in some 
space-like direction $\spd$, $\spd^2=-1$. 
More specifically, we say that a space-time region $\ccc$ {\em contains a 
space-like direction} $\spd$ if 
\begin{equation} \label{eqDirInSpc}
 \ccc+e\subset \ccc \,.
\end{equation}
We take $\Ccc$ to be the set of convex, causally complete regions  which 
contain some space-like direction in this sense.  
(A region $\ccc$ is called causally complete if it contains all points $x$ such
that every inextendible causal curve through $x$ passes through
$\ccc$.)  
Typical examples of regions in $\Ccc$ are space-like cones and {\em wedge} 
regions, {\it i.e.\ }Poincar\'e transforms of the standard wedge 
\begin{align} \label{eqWe}  
W_1 &\doteq\{\,x\in\Bb^3:\,|x^0|<x^1\;\}\,. 
\end{align} 
Wedges are the largest regions in the class $\Ccc$, in the sense that
every $\ccc\in\Ccc$ is contained in some wedge~\cite{BGL}. 

The additional information indicated above, which has to be specified
along with each localization region $\ccc\in\Ccc$, is a path 
in the set  of space-like directions. We denote the latter by 
\begin{equation} \label{eqSpd}
\Spd \doteq \{e\in \Bb^3:e^2=-1\}\,,
\end{equation}
and consider paths in $\Spd$ starting at a reference direction
$\refspd$, which we fix, once and for all, to be 
\begin{equation} \label{eqrefvec}
  \refspd \doteq (0,0,-1)   \,. 
\end{equation} 
Given a region $\ccc\in\Ccc$, we shall say that a path $\spdpath$ {\em ends in}
$\ccc$ if its endpoint is contained in $\ccc$ in the sense of
equation~\eqref{eqDirInSpc}. Two paths $\spdpath_1$ and $\spdpath_2$
starting at $\refspd$ and ending in $\ccc$ will be called {\em
  equivalent w.r.t.\ }$\ccc$ iff the path $\spdpath_1^{-1}\ast\spdpath_2$ 
(the inverse of $\spdpath_1$ followed by $\spdpath_2$) is
fixed-endpoint homotopic to a path which is contained in $\ccc$.  
Now the index set for our localization structure,   
denoted by $\Cccpaths$, is the set of pairs 
\begin{equation} \label{eqCccpath} 
(\ccc,\spdpath)  \,, 
\end{equation}
where $\ccc\in\Ccc$ and 
$\spdpath$ is the equivalence class w.r.t.\
$\ccc$ of a path in $\Spd$ starting at $\refspd$ and ending in $\ccc$. 
{}For fixed $\ccc\in\Ccc$, we shall use the notation $\cccpath$ for an
element of the form $(\ccc,\spdpath)$. 
To see what is involved, suppose $\ccc$ is a space-like cone or a 
wedge. Then the set of directions contained in $\ccc$ is a connected and 
simply connected subset of $\Spd$, and different elements 
$(\ccc,\spdpath_1)$ and  $(\ccc,\spdpath_3)$ differ just by a winding
number, {\it cf.\ }Figure~1.  
\vspace*{2ex}
\begin{figure}[ht] 
 \label{Fig1}
\psfrag{e0}{$e_0$}
\psfrag{e1}{$\spdpath_1$}
\psfrag{e2}{$\spdpath_2$}
\psfrag{e3}{$\spdpath_3$}
\psfrag{C}{$C_H$}
\psfrag{H}{$H$}
\begin{center}
\epsfxsize35ex 
\epsfbox{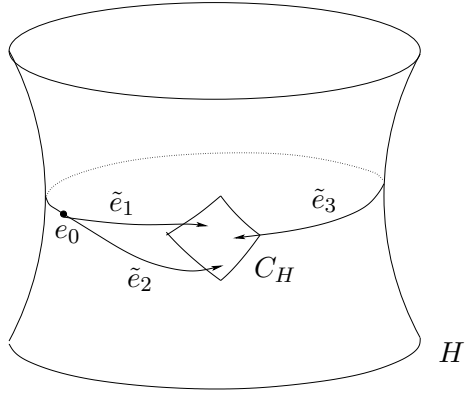}
\caption{$C_H$ denotes the set of space-like directions contained in
  $C$. $(C,\spdpath_1)$ is equal to $(C,\spdpath_2)$, but different
  from $(C,\spdpath_3)$.}
\end{center}
\end{figure}
Consider now two such pairs $\cccpath_1\doteq(\ccc_1,\spdpath_1)$ and 
$\cccpath_2\doteq(\ccc_2,\spdpath_2)$. 
If $\spc_1\subset \spc_2$  and the corresponding paths 
$\spdpath_1$, $\spdpath_2$ are equivalent w.r.t.\ $\spc_2$, then  
we shall write 
\begin{equation} \label{eqSpcSub}
\spcpath_1 \subset \spcpath_2 \,. 
\end{equation}
If $C_1$ and $C_2$ are causally separated, then 
$\cccpath_1$ and $\cccpath_2$ determine a {\em relative winding number} 
\begin{equation} \label{eqWinNum} 
N(\cccpath_1,\cccpath_2)\doteq \text{ winding number of }\quad
\spdpath_2^{-1}\ast \spdpath_1 \ast \spdpath_{12}\;, 
\end{equation}
where $\spdpath_{12}$ is the ``direct'' path from $\spd_1$ to $\spd_2$ in
clockwise direction. 
Finally, we note that the universal covering $\Potild$ of the Poincar\'e group 
naturally acts on $\Cccpaths$ as explained in the appendix, 
{\it cf.\ }equation~\eqref{eqPoincSpc}, such that a $2\pi$ rotation
acts non-trivial --- it maps, for example, $(C,\spdpath_3)$ in 
Figure~1 onto $(C,\spdpath_1)$. 

We now turn to the definition of a localization structure. We admit
the case of several particle species of the same type $(m,s)$, for
example a particle and its anti-particle.  
\begin{Def} \label{DefLS} 
Let $U$ be a finite direct sum of copies of the irreducible
representation of $\Potild$ for mass $m>0$ and spin 
$s\in \Bb$, acting in a Hilbert space $\calH$. 
A family of closed real subspaces $\Ke(\spcpath)$, 
$\spcpath\in\Spcpaths$, 
of $\calH$ is called a {\em localization structure} for $(m,s)$ if it has the 
following properties: 
\begin{liste} 
\item {\em Isotony:} \label{isot}  Let $\spcpath_1 \subset\spcpath_2$
  in the sense of equation~\eqref{eqSpcSub}. Then 
\[ \K(\spcpath_1)\subset\K(\spcpath_2) \;.  \]
\item {\em Twisted Locality:} \label{loc}
There is a complex number $\Z$  of modulus one, such that for any pair 
$\spcpath_1,\spcpath_2\in\Spcpaths$ with $\spc_1$ causally separated 
from $\spc_2$ 
\begin{align}  \label{eqLoc} 
Z(\spcpath_1,\spcpath_2)\,\K(\spcpath_2)\:\subset\:
\K(\spcpath_1)'\;. 
\end{align}
Here, $Z(\spcpath_1,\spcpath_2)\doteq \Z^{2N+1}$, with 
$N=N(\spcpath_1,\spcpath_2)$, cf.~\eqref{eqWinNum}, and the prime 
denotes the symplectic complement\footnote{The relevant notions 
referring to real subspaces of a Hilbert space are recalled in 
Appendix~\ref{secTT}.}.  
\newcounter{App} \setcounter{App}{\thefootnote} 
\item   {\em Poincar\'e covariance:}\label{cov} For all $\spcpath\in\Spcpaths$ 
and $\gtild\in\Potild\,$ 
  \begin{align*}
U(\potild)\:\K(\spcpath)&
 \:= \: \K(\potild\act\spcpath) \;.
  \end{align*}
\item  \label{cycl} 
{\em Standardness:}  $\K(\spcpath)$ is standard$^{\theApp}$ for all 
$\spcpath\in\Spcpaths$. 
\end{liste} 
\end{Def}
\begin{Remark} 
($i$) Covariance implies that 
$K(\rot{2\pi}\act\cccpath)=e^{2\pi is}K(\cccpath)$, where
$\rot{\cdot}$ denotes rotation. Therefore 
$K(\rot{2\pi}\act\cccpath)$ coincides with $K(\cccpath)$ if, and only if,
$s\in\half\Bi$. 
Hence $K(\cccpath)$ is independent of the path $\spdpath$, but only depends
on $\ccc$, iff $s\in\half\Bi$. Further, it can be shown using the free 
field formalism (or along the same lines as in the present 
analysis, {\it cf.\ }the remark after Proposition~\ref{Coc}), that in this case
the localization structure can be extended to bounded regions.   
\\
($ii$) In the framework of algebraic quantum field theory, a field
algebra for anyons~\cite{Re} is a family of operator algebras 
$\{\calF(\cccpath)\}_{\cccpath\in\Cccpaths}$, indexed by the same class 
$\Cccpaths$ (except that in general, each $\ccc$ must contain some 
space-like cone~\cite{BuF}).  Suppose there are finitely many particle
species of the type $(m,s)$ and that $\{m\}$ is isolated from the rest of
the spectrum, and denote by $\Emass$ the projection onto the
corresponding single particle space. Then 
\begin{equation}  \label{eqFsaOm}
K(\cccpath) \doteq \Emass \calF(\cccpath)^{\rm sa}\,\Omega\,^{\clo}  
\quad  \text{ (norm closure)} \,, 
\end{equation}  
$\cccpath\in\Cccpaths$, is a localization structure. This is in fact
the motivation for our definition. 
As an illustration, we show that twisted locality~\eqref{eqLoc} holds in the 
case of bosons or fermions. In these cases field 
operators $\field_1$ and $\field_2$ localized in causally separated regions 
commute or anti-commute, respectively. These relations have been shown 
in~\cite[Sect.\ 2]{BuEp} to survive the projection $\Emass$ in the sense that 
\begin{equation} \label{eqCRAny''} 
\lsp \field_1\Omega,\Emass \field_2\Omega \rsp = \pm 
\lsp \field_2^*\,\Omega,\Emass \field_1^*\,\Omega \rsp\,, 
\end{equation}
respectively. Hence, putting $Z\doteq1$ for bosons and $Z\doteq i$ 
for fermions, the imaginary part of 
$\lsp Z \field_1\Omega,\Emass \field_2\Omega \rsp$ is zero 
if $\field_1$ and $\field_2$ are self-adjoint. This is twisted 
locality. In the general case of anyons, analogous considerations hold,
with $Z$ being defined as a root of the statistics phase. 
\end{Remark}   
We finally recall the definition of a certain maximality property 
called twist\-ed Haag duality. 
Let $\cccpath=(\ccc,\spdpath)$ and $\cccpath'=(\ccc',\spdpath')$,
where $\ccc'$ is the causal complement 
of $\ccc$ and $\spdpath'$ is the equivalence class of a path ending
in $\ccc'$ in the same sense as in equation~\eqref{eqDirInSpc}. 
If $\ccc$ is not a wedge, then the region $\ccc'$ is not contained in
any wedge region. In this case we define a real subspace corresponding to 
$\cccpath'$  via 
\begin{equation} \label{eqKC'} 
K(\spcpath')\doteq 
\bigvee_{\cccpath_0\subset\cccpath', \cccpath_0\in\Cccpaths} K(\cccpath_0) 
\,.   
\end{equation} 
\begin{Def} \label{DefTHD}
A localization structure is said to satisfy {\em twisted Haag
  duality}  if for every pair $\cccpath,\cccpath'$ as above the identity 
\begin{align} \label{eqTHD}
Z(\cccpath,\cccpath')\,K(\spcpath')=  \K(\spcpath)' 
\end{align}
holds. 
\end{Def}
\section{Construction of the Localization Structure} \label{secCon}
Let $U$ be a finite direct sum of copies of the irreducible
representation of $\Potild$ for mass $m>0$ and spin 
$s\in \Bb$. We 
now construct a corresponding localization structure 
along the same lines as in~\cite{BGL}. 

We start with the definition of the localization space associated with
the standard wedge $W_1$, {\it cf.\ }equation~\eqref{eqWe}.
Associated with this  wedge are the Lorentz boosts 
$\lor_1(t)$ leaving $W_1$ invariant and acting on the coordinates $x^0,x^1$ as 
\begin{equation} \label{eqBoox}  
\left( \begin{array}{cc}
 \cosh(t) &  \sinh(t)  \\
 \sinh(t) &  \cosh(t)
 \end{array} \right), 
\end{equation} 
and the reflection $j$ about the edge of $W_1$, 
\begin{equation} \label{eqJ}
j:\,(x^0,x^1,x^2) \mapsto(-x^0,-x^1,x^2)\,. 
\end{equation}
We define  $\Delta$ to be the unique positive operator satisfying 
\begin{align} \label{eqDel}  
\Delta^{it}=U(\lortild_1(-2\pi t))\;, \quad t\in\Bb\,, 
\end{align} 
where $\lortild_1(\cdot)$  denotes the lift of $\lor_1(\cdot)$ to the covering 
group $\Potild$. 
We further pick an anti-unitary involution $J$ satisfying 
\begin{align} 
 JU(\potild)J =U(\jtild \potild \jtild)\,,\quad 
\potild\in\Potild\,,  \label{eqJUgJ}
\end{align}
where $\jtild \cdot \jtild$ denotes the lift of the adjoint 
action\footnote{See~\eqref{eqjtild}.} of $j$ to the covering group 
$\Potild$. Lemma~\ref{JUni} asserts that such an
  involution exists. We mention as an aside, that the localization 
structure which we now construct is independent of the particular 
choice, {\it cf.\ }Proposition~\ref{Unique}. 
We then define a closed operator $S$ by 
\begin{equation}\label{eqPolDec}
\Se\doteq J\,\Delta^{\half}\,.
\end{equation}
This operator is densely defined, antilinear and involutive 
due to the group relation $\jtild \boox{t}\jtild=\boox{t}$, cf.~\cite{BGL}. 
Hence, the eigenspace of $\Se$ for the eigenvalue $1$ is a standard
real subspace, {\it cf.\ }Appendix~\ref{secTT}. We take this subspace 
as our localization space for 
\begin{equation}
  \label{eqWetild}
 \We\doteq(W_1,\spdpath_{W_1})\,, 
\end{equation}
where $\spdpath_{W_1}$ is the equivalence  class of
a path starting from $\refspd$ and staying within $W_1$ (in the sense of
\eqref{eqDirInSpc}); In other words, we put  
\begin{equation} \label{eqKeWe} 
\Ke(\We)\doteq \{\phi\in\dom \,\Se:\;\Se\phi=\phi \}\,. 
\end{equation}
The motivation for this definition will become clear after
Definition~\ref{BiWi}. 
Covariance forces us to define the real subspaces corresponding to 
arbitrary wedges $\Wtild=\potild\cdot\We$ by 
\begin{align}  \label{eqDefKW}
K(\potild\cdot \We)&\doteq U(\potild)\,\Ke(\We) \quad\mbox{ for }
\potild\in \Potild. 
\end{align}
The following lemma asserts that this is well-defined. 
\begin{Lem} \label{KW} 
Let $\potild\in\Potild$ satisfy $\potild\cdot \Wtild_1=\Wtild_1.$ Then 
$U(\potild)\,\Ke(\We)=\Ke(\We).$ 
\end{Lem}
\begin{Proof} 
The set of Poincar\'e transformations $\potild\in\Potild$ which map 
$\Wtild_1$ onto itself is the Abelian group generated by the 
one-parameter subgroups of the translations
along the $2$-axes and of the $1$-boosts $\boox{t}.$ Both
of these subgroups commute with $\jtild$ and with the $1$-boosts,  
hence their representers commute with $\Se,$ which implies the claim. 
\end{Proof}

Next we associate real closed subspaces $K(\cccpath)$ 
to  arbitrary regions $\cccpath\in\Cccpaths$ by intersections: 
\begin{align}  \label{eqKC} 
K(\cccpath)\doteq \bigcap_{\Wtild\supset \cccpath}\,K(\Wtild) \;, 
\end{align}
where the intersection goes over all wedge regions which contain 
$\cccpath$ in the sense of~\eqref{eqSpcSub}.
If $\ccc$ is a wedge, this is consistent with~\eqref{eqDefKW} as a 
consequence of the positivity of the energy~\cite{BGL}. 
Note that if $\ccc$ is not a wedge, then~\eqref{eqKC} is the maximal
subspace one can associate with $\ccc$ in view of locality. 

We now state our main result. 
\begin{Thm} \label{LS}
The family $\{\Ke(\cccpath)\}_{\cccpath\in\Cccpaths}$ constructed
above is a localization structure for $(m,s)$, 
{\it cf.\ }Definition{\rm ~\ref{DefLS}}, with $\Z=e^{i\pi s}$. It 
also satisfies twisted Haag duality, {\it cf.\ }equation~\eqref{eqTHD}.  
Further, the anti-unitary involution $\PCT$  defined by 
$ \PCT \doteq \Z^{-1} J$ is a PCT operator, that is, a 
representer of $\jtild$ in sense of equation~\eqref{eqJUgJ}, which
acts geometrically correctly on the localization
structure:\footnote{The action of $\jtild$ on 
$\cccpath\in\Cccpaths$, denoted $\jtild\act\cccpath$, is explained in 
the appendix, cf.~\eqref{eqJSpc}.}  
\begin{equation} \label{eqUjCorr} 
\PCT\,\Ke(\cccpath)=\Ke(\jtild\act\cccpath)\,,
\quad\cccpath\in\Cccpaths\,.
\end{equation}
\end{Thm} 
It is noteworthy that the ``spin-statistics connection'' $\Z^2=e^{2\pi is}$ 
necessarily holds as a consequence of the definition~\eqref{eqKeWe}, 
as we show in Proposition~\ref{SpiSta} below. 
\begin{Proof}
Isotony and Poincar\'e covariance, i.e.\ properties~\eqref{isot} and 
\eqref{cov} of Definition~\ref{DefLS}, follow
immediately by construction. 
We next prove equation~\eqref{eqUjCorr}. From the group relations
$\boox{t}\jtild=\jtild \boox{t}$, $\boox{t}\rot{\pi}=\rot{\pi}\boox{-t}$ and  
$\rot{\pi}\jtild=\jtild \,\rot{-\pi}$, and the fact that
$\Z^2=e^{2i\pi s}=U(\rot{2\pi})$, it follows that the operator
$U(\rot{\pi})U(\jtild)$ commutes with $\Se.$ But this implies that 
\begin{equation} \label{eqjpiW1}
 U(\jtild)\,\Ke(\We)= U(\rot{-\pi})\,\Ke(\We)=\Ke(\jtild \act\We)\,, 
\end{equation}
where we have used that $\jtild \act\We=\rot{-\pi}\act\We$. Hence, 
equation~\eqref{eqUjCorr} holds for $\cccpath=\We$. By covariance,
it  holds for all wedge regions, and by the intersection 
property~\eqref{eqKC} it holds for all $\cccpath\in\Cccpaths$. 

We next prove twisted Haag
duality~\eqref{eqTHD}. 
Equation~\eqref{eqjpiW1} implies that $J K(\We)=\Z K(\jtild\act\We)$. 
Now according to a  general result about 
Tomita operators, see 
e.g.~\cite[Prop. 2.3]{RvD}, the anti-unitary part 
$J$ in the polar decomposition of $\Se$ maps $K(\We)$ onto its 
symplectic complement: 
\begin{equation} \label{eqJKK'}
J\KeWe = \KeWe'\,.
\end{equation} 
Further, $\Z=Z(\We,\jtild\act\We)$ since the relative winding number
$N(\We,\jtild\act\We)$ is zero. We therefore have 
\begin{equation} 
Z(\Wtild_1,\jtild\act\Wtild_1) \,\Ke(\jtild \act\Wtild_1)= \Ke(\Wtild_1)' \,.
\end{equation}
Now any $\We'=(W_1',\spdpath)$ differs from $\jtild\act \We$ by a
rotation about a multiple of $2\pi$. Replacing $\jtild\act \We$ by
such $\We'$, the above equation is still valid because 
$Z(\We,\rot{2\pi N}\act\jtild\act\We)$ picks up a factor $e^{-2\pi isN}$ which
is compensated by the factor picked up by 
$\Ke(\rot{2\pi N}\act\jtild \act\Wtild_1)$. By covariance and the fact that 
$Z(\potild\act\cccpath_1,\potild\act\cccpath_2)$ is independent of
$\potild\in\Potild$, we get twisted Haag duality  for wedge
regions, i.e. for every pair $\Wtild, \Wtild'$ the identity 
\begin{equation} \label{eqTHDW}
Z(\Wtild,\Wtild') \,\Ke(\Wtild')= \Ke(\Wtild)' 
\end{equation}
holds.  
{}For smaller regions we use a chain of equalities similar to the proof of 
Corollary~3.4 of~\cite{BGL}. Let $\cccpath$ and $\cccpath'$ be as in
Definition~\ref{DefTHD}. Then  
\begin{align} \label{eqTHD'}
Z(\cccpath,\cccpath')\,K(\spcpath')=
Z(\cccpath,\cccpath')\bigvee_{\cccpath_0\subset \cccpath'} K(\cccpath_0)=  
Z(\cccpath,\cccpath')\bigvee_{\Wtild'\subset\cccpath'}
K(\Wtild')&
\nonumber \\
=\bigvee_{\Wtild'\subset\cccpath'}Z(\cccpath,\Wtild')K(\Wtild')=
\bigvee_{\Wtild\supset\cccpath}K(\Wtild)'=
\big(\bigcap_{\Wtild\supset\cccpath}K(\Wtild)\big)'= &K(\cccpath)'\,.
\end{align}
In the second equation we have used the fact that for any pair of 
causally separated regions $\ccc,\ccc_0\in\Ccc$ there is a wedge $W$ such that 
$ \ccc_0\subset W'\subset\ccc'$, cf.~\cite{BGL}, and also that 
\begin{equation} \label{eqKWC} 
  K(\We) = \bigvee_{\cccpath\subset\We} K(\cccpath)\,.  
\end{equation}
This fact is asserted by Takesaki's theorem because the r.h.s.\ is a 
standard space contained in $K(\We)$ and 
is, by equation~\eqref{eqDel} and covariance, invariant under the 
modular group of $K(\We)$. The fourth equation follows from
equation~\eqref{eqTHDW}. We have also used the fact that
$Z(\cccpath_1,\cccpath_2)$ is insensitive to making the regions
$\ccc_1$, $\ccc_2$ smaller. 
We have thus proved twisted Haag duality, which
obviously  implies twisted locality, so we have shown property~\eqref{loc} of 
Definition~\ref{DefLS}. 

It remains to prove property~\eqref{cycl} of the 
Definition~\ref{DefLS}, namely that $\Ke(\cccpath)$ 
is standard for each $\cccpath.$ The real subspace associated to $\We$
(and hence to any other wedge region $\Wtild$) has this property by 
construction, {\it cf.\ }equation~\eqref{eqKeWe} and
Appendix~\ref{secTT}. The property that 
$K(\Wtild)\cap iK(\Wtild)=\{0\}$ transfers to the smaller spaces 
$\Ke(\cccpath)$. It remains to show that  
$\Ke(\cccpath)+i\Ke(\cccpath)$ is dense for all $\cccpath.$ 
But this follows from Corollary~\ref{SFTg} in the next section, bearing
in mind the 
following consequence of the Reeh-Schlieder theorem for the free
scalar massive field: Consider the set of Schwartz functions with compact 
support contained in a fixed  open space-time region. The restrictions 
to the mass shell of the Fourier transforms of these functions are 
dense in the space of square-integrable functions on the mass shell. 
\end{Proof} 

\section{Standardness of the Real Subspaces} \label{secSta}
To prove that $\Ke(\Ctild)+i\Ke(\Ctild)$ is dense, we will explicitly
exhibit sufficiently many elements in $\Ke(\Ctild)$.  
This will be the only place in our analysis where we make explicit use 
of the representation $U$ of $\Potild$. 
It suffices to consider $U$ to be irreducible. For if $U$ is reducible, we
may take the involution $J$, {\it cf.\ }equation~\eqref{eqJUgJ}, as a
direct sum of 
suitable involutions. We then obviously end up with a localization
structure which is the direct sum of irreducible localization
structures. 

We recall the relevant irreducible representations, starting with some 
notational remarks. 
Let $\Lor$ be the Lorentz group in $d=2+1$ and $\Lortild$ its 
universal covering group.\footnote{The relevant 
facts concerning $\Lortild$ and the covering $\Lortild\rightarrow \Lor$ are 
recalled in Appendix~\ref{secPoincUjCorr}.}  
We denote elements of $\Lortild$ generically by $\lortild$, and the 
covering homomorphism $\Lortild\rightarrow \Lor$ by  
\begin{equation} \label{eqCov}
  \lortild \mapsto \lor\,. 
\end{equation}
The group $\Potild$ is the semidirect product of $\Lortild$ with 
the translation group $\Bb^3$. Thus, elements of $\Potild$ will be denoted by 
$\potild=(a,\lortild)$, and the group  multiplication is given  by 
\begin{equation}\label{4.7}
 (a,\lortild)(a',\lortild')=
 (a+\lor a'\,,\, \lortild\lortild').
\end{equation}  
We occasionally denote $(0,\lortild)$ simply by $\lortild$. 
The irreducible representation of $\Potild$ for 
$m>0$ and $s\in\Bb$, denoted by $U$ in the sequel, is given 
as follows. Let $\Hyp$ denote the positive mass shell 
$\{ p\cdot p=m^2, p_0>0\}$ and $d\mu$ the Lorentz invariant measure on 
$\Hyp$. Then $U$ acts on $\calH\doteq L^2(\Hyp,d\mu)$ according to 
\begin{equation} \label{eqUms} 
\big(\U(a,\lortild)\phi\big) (p) = e^{is\Omega(\lortild,p)}\,e^{ia\cdot p}\,
\phi(\lor^{-1}p)\,, 
\end{equation}
where $\Omega(\lortild,p)\in\Bb$ is the Wigner rotation, 
{\it cf.\ }equation~\eqref{eqWigRot}. 
To this representation a unique, up to a phase 
factor, anti-unitary involution $J$ can be adjoined
satisfying equation~\eqref{eqJUgJ}, thus extending $U$ to $\Potildj$ within
the same Hilbert space: 
\begin{equation}
  \label{eqUj}
(J\phi)(p)\doteq e^{i\pi s}\,\overline{\phi(-jp)}\,,
\end{equation}
{\it cf.\ }Lemma~\ref{JUni}. 
Let $\{\K(\cccpath)\}_{\cccpath\in\Cccpaths}$ be the resulting
localization structure as in Theorem~\ref{LS}. 

We now calculate elements in $\K(\cccpath)$ for given $\cccpath\in\Cccpaths$. 
By construction, 
$\phi\in\K(\cccpath)$ if and only if for all $\potild\in \Potild$ which  
map $\cccpath$ into $\We,$\footnote{$\We$ has been defined in
  equation~\eqref{eqWetild}}  the vector $\U(\potild)\phi$ is in 
$\K(\We)$.  
In particular, 
it must be in the domain of $\Delta^\half$. As is well-known~\cite{BiWi}, 
this implies that the map 
\begin{equation} \label{eqUl1g}
 t\mapsto \U(\boox{t})\,\U(\potild)\,\phi\;,\;t\in\Bb , 
\end{equation} 
is the boundary value of an analytic $\calH$-valued function 
on the strip $\Bb+i\,(0,\pi)$. 
But a complication arises from the Wigner rotation factor. 
Namely, the function 
 $t\mapsto \exp (is\Omega(\boox{t}\lortild,p))$  
has singularities in the strip for any fixed $p\in\Hyp$ and 
$\lortild\in\Lortild$ in a neighbourhood of the unit, which are  
branch points if $s$ is not an integer (see Lemma~\ref{Oml1}). 
Our strategy is to 
consider wave functions of the form $\phi=u\cdot\psi$ (point-wise
multiplication),  
where $u$ is a fixed non-vanishing function on the mass shell, 
suitably chosen as to compensate the singularities of the Wigner 
rotation factor. The action of $\U(\potild)$, according to 
equation~\eqref{eqUms}, on wave functions of the form 
$(u\cdot\psi)(p)\doteq u(p) \psi(p)$ can be written as    
\begin{align}
\big(\U(a,\lortild)\, u\cdot\psi\big) (p)&= 
u(p) \,\coc(\lortild,p)\;e^{i a\cdot p}\; 
\psi(\lor^{-1}p)\,, \label{eqUCoc} 
\intertext{ with }  
\coc(\lortild,p)&\doteq u(p)^{-1}\;e^{is\Omega(\lortild,p)}
\;u(\lortild^{-1}\,p) \,.\label{DefCoc}
\end{align}
In group theoretical terms, the map 
$c(\cdot,\cdot):\Lortild\times\Hyp\rightarrow \Bc\setminus \{0\}$ 
is a cocycle
which is equivalent to the Wigner rotation factor.  
As indicated above, our strategy is to choose $u$ such that
$c(\lortild,p)$ has the desired analyticity properties. 
This will succeed only for certain $\lortild\in\Lortild$
or, differently stated, for certain $\cccpath\in\Cccpaths$. We shall
consider, as a first step, $\cccpath$ of the form $(\ccc,\refspdpath)$, 
with  $\ccc$  containing the reference direction $\refspd$, cf.\
equation~\eqref{eqrefvec}, and where 
$\refspdpath$ denotes the constant path at $\refspd$. 
Stated differently, we consider elements $\lortild\in\Lortild$ which 
satisfy 
\begin{equation}
  \label{eqDirInSpcPath}
 \lortild \act \refspdpath \in \We\,. 
\end{equation}
By this we mean that $W_1$ contains the direction $\lor\act \refspd$
in the sense of equation~\eqref{eqDirInSpc}, and that the paths
  $\lortild\act\refspdpath$ and $\spdpath_{W_1}$, cf.\
  equation~\eqref{eqWetild},  are equivalent   w.r.t.\   $W_1$. 
The following function is suitable for this purpose, and in the 
sequel the cocycle $\coc$ will be defined as in
equation~\eqref{DefCoc} above with this choice of $u$:  
\begin{align} \label{equ} 
  u(p)&\doteq \Big(\frac{p_0-p_1}{m}\cdot
\frac{{p_0-p_1+m-ip_2}}{p_0-p_1+m+ip_2}\Big)^{s}\;,\quad 
p_0\doteq ({p_1^2+p_2^2+m^2})^{\half}\,.
\end{align}
Note that $p_0-p_1$ is strictly positive for all $p\in\Hyp,$ hence the 
argument in brackets lies in the cut complex plane 
$\Bc\setminus \Bb^-_0.$ Thus, taking it to the power of $s\in\Bb$ can
be defined via the
branch of the logarithm on $\Bc\setminus \Bb^-_0$ with $\ln 1=0.$  
This will always be understood in the sequel and will 
be called the power of $s$  {\em within} $\Bc\setminus \Bb^-_0$.  
\begin{Lem} \label{CocW1}
Let 
$\lortild$ be an element of $\Lortild$ such that 
$\lortild\act\refspdpath\in\We$ in the sense of 
equation~\eqref{eqDirInSpcPath}.  
Then for all $p\in\Hyp$ the function   
$$t\mapsto \coc(\boox{t}\lortild,p) 
$$
has an analytic extension into the strip $\Bb+i(0,\pi)$.  
This extension satisfies the boundary condition 
\begin{equation} \label{eqCociPi}
 \coc(\boox{t+i\pi}\lortild,p)=e^{i\pi s}\,
\overline{\coc(\boox{t}\lortild,-jp)}\,, \quad t\in\Bb.  
\end{equation}
\end{Lem}
\begin{Proof}
As we show in Lemma~\ref{Cover}, $\lortild$ can be decomposed into
boosts and rotations as $\lortild=\boox{t}\;\boo{2}{t'}\;\rot{\omega}$
for some unique $t,t',\omega\in\Bb.$ We then denote 
$\omega'\doteq\omega-\pihalf$. Then $\lortild\act\refspdpath\in\We$ if 
and only if 
\begin{equation} \label{eqLorW1}
\Boo{2}{t'}\,\Rot{\omega'}\,(0,1,0) \in W_1^{\clo} \quad\text{ and } \quad
\omega'\in (-\pi,\pi)\,, 
\end{equation}
the latter condition singling out the 
correct leaf of the covering $\rot{\omega'}\mapsto\Rot{\omega'}$. 
As the vector in equation~\eqref{eqLorW1} is equal to 
$$(\sinh t'\,\sin\omega',\cos\omega',\cosh t'\,\sin\omega' ),$$
condition~\eqref{eqLorW1} is equivalent to 
\begin{equation} \label{eqLorW1'}
|\sinh t'\,\sin\omega'|
\leq \cos\omega' \quad \text{ and }\quad  \omega'\in [-\pihalf,\pihalf]\,. 
\end{equation}
This implies condition~\eqref{eqCocAna} of Proposition~\ref{Coc} in
Appendix~\ref{proof}, 
which now  asserts the claimed analyticity property and 
the correct boundary value of the cocycle. 
\end{Proof}

We denote by $C_0^\infty(\Bb^3)$ the $C^\infty$-functions on $\Bb^3$
with compact support, and, for $f\in C_0^\infty(\Bb^3)$,  by $E_mf$ the 
restriction of the Fourier transform of $f$ to the mass shell $\Hyp.$ 
Our main result is the following proposition.  
\begin{Prop} \label{SFT} 
Let $\ccc$ be a region in $\Ccc$ containing the reference direction
$\refspd$  in the sense of equation~\eqref{eqDirInSpc}, 
and let $\cccpath=(\spc,\refspdpath).$ Then 
\[ \K(\cccpath)\;\supset\;
\{ u\cdot E_m{f}\;|\; {f}\in C_0^\infty(\spc), \;\text{ real valued } 
\;\}\,.
\]
\end{Prop}
Before proving the proposition, we point out that the local subspaces
for regions containing directions other than
$\refspd$ are obtained via covariance, and can be nicely characterized
as follows. Define, for each $\lortild\in\Lortild$, a function 
$u_\lortild$ on the mass shell by 
\begin{equation} \label{equLor}
 u_\lortild(p) \doteq u(p)\,\coc(\lortild,p)\,. 
\end{equation}
This is an ``intertwiner function'' for those single particle vectors
which are localized in regions extending to infinity in the direction 
$\lortild\act\refspdpath$: 
\begin{Cor}  \label{SFTg}
i) 
Let $\lortild \in\Lortild$ and $\cccpath\in\Cccpaths$. If $\cccpath$
contains $\lortild\act\refspdpath$ in the sense of
equation~\eqref{eqDirInSpcPath}, then 
\[ \K(\cccpath)\;\supset\;
\{ u_\lortild \cdot E_m{f}\;|\; {f}\in C_0^\infty(\spc),\;\text{ real valued } 
\;\}\,.
\]
ii) The wave functions $ u_\lortild \cdot E_m{f}$ transform
covariantly in  the sense that 
\begin{equation}
  \U(a,\lortild)\,u_{\lortild'}\cdot E_mf =
  u_{\lortild\lortild'}\cdot E_m(a,\lor)_*f \,,
\end{equation}
where the star denotes the push-forward, $(g_*f)(x)\doteq f(g^{-1}x)$. 
\end{Cor}
\begin{Proof}
 $i)$ is an immediate consequence of Proposition~\ref{SFT}, and $ii)$
 follows from the cocycle relation~\eqref{eqCocRel} below. 
\end{Proof}   

It is noteworthy that the function $u_\lortild$ only depends on the
path $\lortild\act\refspdpath$ up to a multiplicative constant. For
the stabilizer subgroup of $\refspdpath$, namely the group of 1-boosts,
modifies $u_\lortild$ only by a factor $\coc(\boox{t},p)=e^{st}$.  
\begin{Proofof}of Proposition~\ref{SFT}.  
Let $f$ be a smooth function with compact support in $\spc$, and let 
$\potild=(a,\lortild)$ be such that $\potild \cccpath\subset \We$. 
Note that then $\lortild \refspdpath\in \We$ and $\supp\, g_*f\subset
W_1$, where $\po_*f$ denotes the push-forward as above. 
We have to show that $U(\potild)\, u\cdot E_mf\in\K(\We)$. 
To this end we prove that the $\calH$-valued function 
\begin{equation} \label{eqPhit}
 t\mapsto\phi(t) \doteq \U(\boox{t})U(\potild)\, u\cdot E_mf \;,\quad t\in\Bb,
\end{equation}
is the boundary value of an analytic function $\phi(\cdot)$ on the strip 
$\strip\doteq \Bb+i\,(0,\pi)$ which is continuous and bounded on its
closure $\strip^{\clo}$ 
and that the boundary values are related by 
\begin{equation} \label{eqPiJ}
\phi(t+i\pi) =  J\, \phi(t)\,,\quad t\in\Bb\,. 
\end{equation}
Using the push-forward to write 
$e^{ia\cdot p} (E_mf)(\lor^{-1}p) = \big(E_m(a,\lor)_*f\big)(p)$, 
we have 
\begin{align}
\phi(t)&= 
\U(\boox{t}a,\boox{t}\lortild) \,u\cdot E_mf = v(t)\cdot \psi(t)  \\
\intertext{where we have written }
v(t)(p)&\doteq u(p)\,\coc(\boox{t}\lortild,p) \;,\\
\psi(t)(p)&\doteq \big(E_m\Boox{t}_* \po_*f \big)(p)\,. 
\end{align}
It follows from Lemma~\ref{CocW1} that for fixed $p\in\Hyp$, 
$v(\cdot)(p)$ extends to an analytic function $v(\cdot,p)$ on the strip 
$\strip,$ continuous on its closure, and that 
\begin{equation} \label{eqvipi}
v(t+i\pi,p)= e^{i\pi s} u(p)\,\overline{\coc (\boox{t}\lortild,-j p)}\,.
\end{equation}
Let us discuss the analyticity properties of $\psi(t)$. The matrix-valued 
function $t\mapsto \Boox{t}$ extends to an entire analytic function satisfying 
\begin{equation} \label{eqBooCom}
 \Boox{t+it'} = \Boox{t}\big(j_{t'}+i \sin(t')\,\sigma \big) \,,
\end{equation}
where $j_{t'}$ acts as multiplication by $\cos t'$ on the coordinates 
$x^0$ and $x^1$ and leaves the other coordinates unchanged, and
$\sigma$ acts as the Pauli matrix $\sigma_1$ on $(x^0,x^1)$ and as 
zero on $x^2$~\cite{H96}. 
Hence $\psi(\cdot)(p)$ extends, for fixed $p\in\Hyp,$ to a function
$\psi(\cdot,p)$ on $\strip^\clo$  as follows:  
\begin{equation} \label{eqPsit}
  \psi(t+it',p) \doteq (2\pi)^{-3/2}\,\int_{W_1} \d^3 x \; 
 e^{i\,p\cdot\Boox{t}j_{t'}x}\,e^{-\sin t'\,p\cdot\Boox{t}\sigma x} 
\;(\po_*f)(x)\;.
\end{equation}
Now for $x\in W_1,$ 
the vector $\sigma x$ lies in the forward light cone, hence $p\cdot\Boox{t}
\sigma x >0$ for $p\in\Hyp.$ Thus the second exponential term in
equation~\eqref{eqPsit} is a damping factor, and for fixed
$\tau\in\strip^\clo$ the function
$(p_1,p_2)\mapsto\psi(\tau,p)$ is of fast decrease. 
Further, due to the damping factor the function $\tau \mapsto
\psi(\tau,p)$ is analytic on the strip $\strip$ for fixed $p\in\Hyp$.  
Thus our function $t\mapsto \phi(t)= v(t) \psi(t)$ extends, point-wise in
$p,$ to a function $\phi(\tau,p)\doteq v(\tau,p)\psi(\tau,p)$ on 
$\strip^\clo$, analytic on the
interior, and in addition $\phi(\tau,\cdot)\in L^2(\Hyp,d\mu)$ for
each $\tau\in\strip^\clo$.  
By equations~\eqref{eqvipi} and \eqref{eqBooCom} the analytic continuation 
satisfies, since $j_{\pi}=j,$ 
\begin{equation}  \label{eqPhiipi}
  \phi(t+i\pi,p) = e^{i\pi s} u(p) \,
\overline{\coc (\boox{t}\lortild,-j p)}\,\big(E_m j_*g_*f\big) (p)\,.
\end{equation}
On the other hand, using $u(-jp)=\overline{u(p)}$ one calculates 
\begin{align}
\big(J\, \phi(t)\big)(p)&= e^{i\pi s}
u(p)\, \overline{\coc(\boox{t}\lortild,-jp)}\, 
\big(E_mj_*\po_*\bar{f}\big)(p)\;. 
\end{align}
Thus for real valued $f,$ the Hilbert space valued function
$\tau\mapsto \phi(\tau)$ defined by $\phi(\tau)(p)\doteq \phi(\tau,p)$ 
satisfies the desired equation~\eqref{eqPiJ}. It remains
to show that  $\phi(\tau)$ is in fact analytic as
a Hilbert space valued function. 

To this end let, for $x\in W_1,$ $t_x\doteq {\rm artanh}\frac{x_0}{x_1}.$
Then $\sigma x = |\sigma x|\,\Boox{t_x}(1,0,0)$ and 
\begin{equation} \label{eqpsigx}
 p\cdot\Boox{t}\sigma x = |\sigma
 x|\,\big\{\cosh(t+t_x)p_0-\sinh(t+t_x)p_1
\big\} \,.
\end{equation}
Note that the argument in curly brackets is strictly larger than
$|p_2|$ and than $|p_1|\times$ $\exp(-|t+t_x|)$.
Let $t+it'$ be contained in some compact subset $G_0\subset\strip$ 
of the strip. Then 
\begin{equation}  \label{eqTeps}
 |t| \leq T \quad \text{ and } \sin t' \geq \eps \quad \text{ for some  } 
T>0, \eps >0\;.   
\end{equation}
Then the above estimates imply, using that 
$\exp(-|t_x|) =\big(\frac{x_1+|x_0|}{x_1-|x_0|}\big)^{\half}$, that 
\begin{align}  \label{eqEstim}
 \sin t'\,p\cdot\Boox{t}\sigma x &> \alpha_1(x)|p_1|+\alpha_2(x)|p_2|
 \quad\text{ where }  \\
\alpha_1(x) &\doteq \frac{\eps}{2} e^{-T}(x_1-|x_0|)>0\,,\\
\alpha_2(x)&\doteq \frac{\eps}{2} (x_1^2-x_0^2)^\half >0\,.
\end{align}
This estimate implies that 
\begin{equation} \label{eqDomFct}
 \Psi(p_1,p_2) \doteq \int_{W_1} d^3x \,|g_*f(x)| \,
 e^{-\alpha_1(x)|p_1| -\alpha_2(x)|p_2|}  
\end{equation}
is a dominating function for $\psi(\tau,\cdot)$ for all $\tau$ in the compact 
subset $G_0$ of the strip, in the sense that
$|\psi(\tau,p)|<\Psi(p_1,p_2)$ for all $\tau\in G_0.$  
This function is decreasing fast enough such that 
\begin{equation} \label{eqDomPol}
 \int d^2p \,|p_1|^n|p_2|^m\,|\Psi(p_1,p_2)|^2 \;<\infty\quad\text {for
   all }\;n,m\in\Bn_0\,. 
\end{equation}
Namely, the integral coincides with $4n!m!$ times the integral of 
$|g_*f(x)g_*f(y)|$ $(\alpha_1(x)+\alpha_1(y))^{-n-1}$ 
$(\alpha_2(x)+\alpha_2(y))^{-m-1}$ over $x$ and $y $ in $W_1$, 
which is finite since
$\alpha_1,\alpha_2$ are strictly positive functions on $W_1$ 
and $\supp (g_*f)$ is compactly contained in $W_1$. 
By similar considerations one gets a dominating function for
$\frac{\rm d}{{\rm d} \tau}\psi(\tau,p)$, which we denote by $\Psi'$
and which satisfies the analogue of equation~\eqref{eqDomPol}.  

Next we establish bounds for $v(\tau,p)$: We claim that $v(\tau,p)$
and $\frac{{\rm d}}{{\rm d}\tau}v(\tau,p)$ are bounded, 
uniformly in $\tau\in G_0$, by polynomials in $|p_1|$ and $|p_2|$ which
we denote by $V$ and $V'$, respectively. 
We demonstrate here the case of non-negative 
spin $s,$ the other case working analogously. 
One has the inequality 
$0< p_0\pm p_1 \leq 2|p_1|+|p_2|+m$ and, using the identity 
$\frac{-p_2+im}{p_0-p_1} =i\,\frac{p_0+p_1+m+ip_2}{p_0-p_1+m-ip_2}$, 
the inequality  
$|(-p_2+im)(p_0-p_1)^{-1}|\leq 2(|p_1|+|p_2|+m)/m$. These imply, for 
$\tau \in G_0$, the estimate 
\begin{equation*} 
 v(\tau,p) \leq c_0 (2|p_1|+|p_2|+m)^n\big(
 c_1+ c_2(|p_1|+|p_2|+m)\big)^{2n}\,\doteq
 V(|p_1|,|p_2|)\,, 
\end{equation*}
where $n$ is any integer $\geq s$, $c_1=|a-b|$ and 
$c_2=\frac{2}{m}e^{T-t}|a+b|$ with  $a$ and $b$ as in Proposition~\ref{Coc}. 
Similar considerations hold for $s<0,$ 
and for $\frac{{\rm d}}{{\rm d}\tau}v(\tau,p)$.    

We have now established the following facts: $\phi(t)$ extends to a family 
$\phi(\tau)\in L^2(\Hyp,d\mu),$ $\tau\in \strip^\clo$, such that 
$\phi(\tau)(p)$ depends analytically on $\tau$ for each
$p\in\Hyp.$ Further, for $\tau$ in any fixed compact subset of the
strip $\strip$, the
$p$-point-wise derivative w.r.t.\ $\tau$ is dominated
by a function $\Phi\in L^2(\Hyp,d\mu)$: 
\begin{equation} \label{eqPhiDom}
 |\frac{{\rm d}}{{\rm d}\tau}\phi(\tau)(p)|\leq \Phi(p)\doteq 
V(\bfp)\Psi'(\bfp) + V'(\bfp)\Psi(\bfp)\;,\quad p=(\omega(\bfp),\bfp)\,.  
\end{equation}
That $\Phi$ is in $L^2(\Hyp,d\mu)$ follows from
equation~\eqref{eqDomPol} and the corresponding equation for $\Psi'$. 

These facts imply, by the Lebesgue lemma on dominated convergence,
that for arbitrary $\chi\in L^2(\Hyp,d\mu)$, the function 
$$ \tau \mapsto \big(\chi,\phi(\tau)\big)  
$$ 
is analytic on the strip $\strip$, with derivative being
calculated via the $p$-point-wise derivative 
$\frac{{\rm d}}{{\rm d}\tau}\phi(\tau)(p)$. Since weak and strong analyticity
are equivalent, this implies that $\tau\mapsto \phi(\tau)$ is an
analytic Hilbert space valued function. This concludes the proof. 
\end{Proofof} 
\section{Implications of the Bisognano-Wichmann Property} \label{UniSpiSta}
In this section we show that the \BiWi\!, defined below, essentially fixes the 
localization structure, and that it implies the spin-statistics connection 
as mentioned after Theorem~\ref{LS}.

Given a localization structure 
$\Ke(\spcpath)$, $\spcpath\in\Spcpaths$, denote by $\Se$ the
canonical involution corresponding to $\Ke(\We)$, 
{\it cf.\ }Appendix~\ref{secTT}. Since
$\Se$ is a closed antilinear involution, it has a polar decomposition
$\Se=:J\Delta^{1/2}$ with $J$ being an anti-unitary involution 
and $\Delta$ a positive operator.  
\begin{Def}  \label{BiWi} 
A localization structure satisfies the {\em \BiWi} if $\Delta^{it}$
and $J$ satisfy equations~\eqref{eqDel} and \eqref{eqJUgJ}, 
thus representing the boosts and the reflection $\jtild$, respectively. 
\end{Def}
It is noteworthy that this property in fact {\em follows} from the
Definition~\ref{DefLS} of a localization structure. This has been
established  by the
author in~\cite{M01a} in the case of four-dimensional theories, and 
will be published elsewhere for anyons in d=3. Because of this fact 
we have been forced to take equations~\eqref{eqDel} to \eqref{eqKeWe} as 
the starting point of our construction. 
 
We shall now see that the \BiWi fixes uniquely a certain extension 
of the localization structure 
which is maximal in the sense that it
satisfies twisted Haag duality, {\it cf.\ }\eqref{eqTHD}. 
\begin{Prop} \label{Unique} 
There is up to equivalence only one localization structure which 
satisfies the \BiWi and twisted Haag duality. 
\end{Prop} 
By equivalent localization structure we mean a family 
$\hat{K}(\cccpath)$, $\cccpath\in\Cccpaths$, of closed real subspaces 
of a Hilbert space $\hat{\calH}$ such that there is a unitary map 
$V:\calH\rightarrow \hat{\calH}$ satisfying
$\hat{K}(\cccpath)=V\,K(\cccpath)$ for all $\cccpath\in\Cccpaths$. 
\begin{Proof}
Let 
$K(\cccpath)$, $\cccpath\in\Cccpaths$, be a  localization structure 
as in the Proposition. With the same argument as in the proof of
Theorem~\ref{LS}, 
equation~\eqref{eqKWC} must hold for $ K(\We)$. Hence, 
the chain of equations~\eqref{eqTHD'} is valid, the last equation of
which shows that, under the assumption of twisted Haag duality, 
$ K(\cccpath)$ is maximal in the sense that it satisfies 
equation~\eqref{eqKC}. But this implies that the localization
structure is fixed by the real subspaces associated to wedge regions, 
which in turn are fixed, due to the \BiWi and covariance, 
by the real subspace $ K(\We)$ associated to
$\We$ and the representation $U$. 
Hence the localization structure is fixed by $ K(\We)$ or, equivalently,
by the corresponding involution $ S$. The positive part of the latter 
is fixed by the representation $U$, {\it cf.\ }equation~\eqref{eqDel}, hence
the only remaining freedom is the anti-unitary part $J$. 
But it turns out that $J$, and hence the entire localization 
structure, is fixed up to equivalence. More precisely, 
let $\hat{K}(\cccpath)$, $\cccpath\in\Cccpaths$, be another localization 
structure as in the Proposition, with $\hat J$ the anti-unitary
part of the canonical involution corresponding to $\hat{K}(\We)$.  
Then, as we show in Lemma~\ref{JUni}, there is a unitary $V$ commuting
with the representation $U$ such that $\hat{J}=VJV^{-1}$. 
This implies that $\hat{K}(\cccpath) = V {K}(\cccpath)$ for all
$\ccc\in\Cccpaths$, as claimed. 
\end{Proof}

We finally prove a single-particle version of the spin-statistics
theorem: 
\begin{Prop} \label{SpiSta}
Let $\{K(\cccpath)\}_{\cccpath\in\Cccpaths}$ be a localization
structure for $(m,s)$ satisfying the \BiWi\!. Then 
the spin-statistics connection holds: 
\begin{equation} \label{eqSpiSta}
  \Z^2 = e^{2 \pi is} \,. 
\end{equation}
\end{Prop}
\begin{Proof}
We use the one-to-one correspondence between closed real
standard subspaces $K$ and densely defined anti-linear involutive 
operators $S$, {\it cf.\ }Appendix~\ref{secTT}.   
Let $S'$ be the canonical involution corresponding to $K(\rot{\pi}\We)$. 
Twisted locality~\eqref{eqLoc} implies that 
$$Z(\We,\rot{\pi}\We)\, S'\,Z(\We,\rot{\pi}\We)^*\subset\Se^*.$$ 
Now the relative winding number
$N(\We,\rot{\pi}\We)$ is $-1$, hence 
$$Z(\We,\rot{\pi}\We)=\Z^{-1}$$ 
and we have $S'$ $\subset$ $\Z^{2}$ $\Se^*$. 
On the other hand, 
$$S'=U(\rot{\pi})\,\Se \,U(\rot{-\pi})$$ 
by covariance. But the group relations imply~\cite{BGL} that $\Se$ 
$\,U(\rot{-\pi})$ $=U(\rot{\pi})$ $\,\Se^*,$ hence
$\Z^{2}=U(\rot{2\pi})$ $\equiv e^{2\pi i s}$, which proves the claim.
\end{Proof}

\appendix
\section{Basic Notions from the Tomita-Takesaki Theory of Real Spaces}
\label{secTT} 
{}For a review of this theory, the reader is referred to one of the articles 
\cite{LRT,RvD,BGL}.  Here we recall the relevant notions. 

Let $\calH$ be a (complex) Hilbert space with scalar product
$\lsp  \cdot\,,\cdot \rsp$. If $K$ is a real subspace of $\calH$, then its 
{\em symplectic complement} is the set of vectors $\psi\in\calH$ such 
that the imaginary part of $\lsp \phi,\psi \rsp$ vanishes for all 
$\phi\in K$.  It is a closed real subspace and is denoted by $K'$.  
If $K_\alpha$, $\alpha \in I$, is a family of closed real subspaces,
then the closed real span of these subspaces is 
denoted by $\bigvee_{\alpha\in I}K_\alpha$.  Its symplectic complement
is given by $(\bigvee_{\alpha\in I}K_\alpha)'=\bigcap_{\alpha\in I}K_\alpha'$. 

A real closed subspace $K$ of $\calH$ is called {\em
  standard} if $K+iK$ is dense in $\calH$ and $K\cap iK=\{0\}$. 
Real closed standard subspaces $K$ of $\calH$  are in one-to-one 
correspondence with antilinear, densely defined, closed operators $S$ 
acting on $\calH$ which are involutive ({\it i.e.}, satisfy 
$S^2\subset \unity$): Given $S$, let 
\begin{equation} \label{eqK1}  
 K \doteq \{\,\phi\in \dom S:\,S\,\phi=\phi\,\} \;. 
\end{equation}
Then every vector in the domain of $S$ may be uniquely written as
$\psi=\phi_1+i\phi_2$ with $\phi_1,\phi_2\in K$, 
namely $\phi_1\doteq\half(\psi+S\psi)$ and $\phi_2\doteq \frac{1}{2i}
(\psi-S\psi)$.  Hence $K$ is standard. 
It is called the real space corresponding to $S$.  
Conversely, a real closed standard subspace $K$ defines an antilinear, 
densely defined, closed involution $S$, by putting $S(\phi_1+i\phi_2)\doteq
\phi_1-i\phi_2$ for $\phi_1,\phi_2\in K$. $S$ is then called the {\em 
canonical involution} corresponding to $K$.   
If $S$ corresponds to $K$ and $U$ is unitary, then $USU^*$ 
corresponds to $UK$, and further $S^*$ corresponds to $K'$.   
\section{The Universal Covering Group of the \\ Poin\-car\'e Group}  
\label{secPoincUjCorr}
\paragraph{Covering of the Lorentz group.} 
The universal covering group $\Lortild$ of the proper orthochronous
Lorentz group $\Lor$ in three dimensions can be identified with the set
\begin{equation}\label{4.4}
 \big\{(\gamma,\omega)\bigm|\gamma\in\Bc,\,|\gamma|<1,\,\omega\in\Bb\big\}\,,
\end{equation}
the group multiplication $(\gamma,\omega)(\gamma',\omega')=(\gamma'',
\omega'')$ being given by~\cite[p.\ 594]{Ba1}
\begin{eqnarray}\label{4.5}
 \gamma''&=&(\gamma'+\gamma e^{-i\omega'})(1+\gamma\bar{\gamma'}e^{-i\omega'})
  ^{-1} \\
 \omega''&=&\omega+\omega'+\einsi \log\Big\{\,(1+\gamma\bar{\gamma'}
e^{-i\omega'})
 (\text{c.c.})^{-1}\,\Big\}\;. \nonumber
\end{eqnarray}
Here (c.c.) denotes the complex conjugate of the preceding
factor and $\log$ is the branch of the logarithm on 
$\Bc\setminus \Bb_0^-$ with $\log 1=0$. 
 
The covering homomorphism $\Lortild\rightarrow\Lor$ is conveniently
described via the double covering $SU(1,1)$ of $\Lor$, which is the 
subgroup of $SL(2,\Bc)$ (conjugate to $SL(2,\Bb)$) consisting of elements 
of the form 
\begin{equation} \label{4.1}
 \left(\begin{array}{cc} \alpha &\beta\\ \bar{\beta}&\bar{\alpha} 
 \end{array}\right)\,, \quad
 \alpha\bar{\alpha}-\beta\bar{\beta}=1.
\end{equation} 
The covering homomorphism $\Lortild\rightarrow SU(1,1)$ associates to
each $(\gamma,\omega)$ the $SU(1,1)$-matrix 
\begin{equation}\label{4.6}
 (1-|\gamma|^2)^{-\half}\;\left(\begin{array}{cc}
 e^{-i\frac{\omega}{2}}&\bar{\gamma} e^{-i\frac{\omega}{2}} \\
 {\gamma}e^{i\frac{\omega}{2}} & e^{i\frac{\omega}{2}}
 \end{array}\right).
\end{equation}
The double covering $SU(1,1)\rightarrow\Lor$ is given as follows. 
{}For $a=(a^0,a^1,a^2)\in\Bb^{3}$  we set 
\begin{equation} \label{4.2a} 
 \utilde{a}=\left(\begin{array}{cc} a^0&a^1-ia^2\\a^1+ia^2& a^0
   \end{array} \right)\,.
\end{equation} 
Then the double covering $SU(1,1)\rightarrow\Lor$ associates to
$A\in SU(1,1)$ the unique $\lambda\in\Lor$ satisfying 
\begin{equation}\label{4.2b}
 \utilde{\lambda a}=A\utilde{a}A^*\, ,\quad a\in\Bb^3.
\end{equation}

Let us determine the lifts of the one-parameter subgroups of boosts and 
rotations. Denote the boosts in 
$k$-direction ($k=1,2$) by $\Boo{k}{\cdot}$ and the rotations in the 
$1$-$2$ plane by $\Rot{\cdot}$. Explicitely, $\Boo{k}{t}$ acts on the $0$- and 
$k$-coordinates as the matrix~\eqref{eqBoox}, 
and $\Rot{\omega}$ acts on the $1$- and $2$-coordinates as  
\begin{equation} \label{eqRot}  
\left( \begin{array}{cc}
 \cos(\omega) &  -\sin(\omega)  \\
 \sin(\omega) &  \cos(\omega)
 \end{array} \right). 
\end{equation}
We denote by 
$\boox{\cdot}\,,\; \boo{2}{\cdot}\;\text{ and }\; \rot{\cdot}$ 
the unique lifts of these one-parameter groups to $\Lortild.$ 
\begin{Lem} \label{Cover}
i) The lifts of the one-parameter groups are given by 
\begin{align} 
\boox{t}= (\tanh({t}/{2}),0) \;,\; 
\boo{2}{t} = (i\tanh({t}/{2}),0) \quad
 \text{ and }\quad  \rot{\omega} = (0,\omega).   
\end{align}
ii) Every element $\lortild\in\Lortild$ has a unique decomposition 
\begin{equation} \label{eql1l2r''} 
\lortild=\boox{t}\;\boo{2}{t'}\;\rot{\omega}\;\quad t,t',\omega\in\Bb\,. 
\end{equation} 
\end{Lem}
\begin{Proof}
$i)$ One verifies that the three one-parameter maps are continuous and
have the correct images under the covering projection~\eqref{4.6} and 
\eqref{4.2b}. $ii)$ 
Consider the action of 
the Lorentz transformation $\lor$, corresponding to
$\lortild$,  on the point $(1,0,0)$. 
Define $t'$ as the ${\rm arsinh}$ of the $2$-component of 
$\lor\act(1,0,0)$, and $t$ as the unique real number such that
$\sinh(t)\cosh(t')$ is the $1$-component of $\lor\act(1,0,0)$. 
One then checks that the actions of $\lor$ and $\Boox{t}\Boo{2}{t'}$ on
the point $(1,0,0)$ coincide. 
This implies that there is a unique $\omega\in\Bb$ such that
equation~\eqref{eql1l2r''} holds. 
\end{Proof}
\paragraph{Wigner Rotation.} 
Let, for $p\in\Hyp,$ 
\begin{equation} \label{eqgamp} 
 \gamma(p)\doteq \frac{p_1+ip_2}{p_0+m}\;,\quad  
\tilde{h}(p)\doteq\big(\gamma(p)\,,\,0\big)\,,  
\end{equation}
and denote by $h(p)$ the corresponding element in $\Lor.$ Then 
\begin{equation} \label{eqBoop}
 h(p) :(m,0,0) \mapsto p\,. 
\end{equation}
This implies that for arbitrary $p\in\Hyp$ and 
$\lortild\in\tilde{L}_3^\uparrow$, the element
\begin{equation}\label{4.8a}
 t(\lortild,p)\doteq\tilde{h}(p)^{-1}\,
\lortild\;\tilde{h}(\lor^{-1}p)
\end{equation} 
leaves $(m,0,0)$ invariant, hence is a rotation and may be written 
in the form
\begin{equation}\label{4.8b}
 t(\lortild,p)=
 \big(0,\Omega(\lortild,p)\big),
\end{equation} 
where $\Omega(\cdot,\cdot)$ is the so-called Wigner rotation. 
In fact, equations (\ref{4.8a}) and (\ref{4.5}) imply that, for
$\lortild=(\gamma,\omega)$, 
\begin{align}\label{eqWigRot}
\Omega(\lortild,p)=
 \omega&+\einsi\log\Big\{\big(1-\gamma(p)\bar{\gamma}e^{-i\omega}\big)
\big({\text{c.c.}}\big)^{-1}\Big\}\nonumber\\
 &+\einsi\log\left\{\Big(1+\frac{\gamma-\gamma(p)e^{-i\omega}}
 {1-\gamma(p)\bar{\gamma}e^{-i\omega}}
        \bar{\gamma}
  (\lor^{-1}p)\Big)\,\Big(\text{c.c.}\Big)^{-1}\right\}\;.
\end{align}
Note that $\Omega\big((0,\omega)\,,\,p\big)=\omega$ for all
$\omega$ and $p$, and that $\Omega$ satisfies the cocycle condition
\begin{align} \label{eqOmCoc}
 \Omega(\lortild\lortild',p)=
\Omega(\lortild,p)+\Omega(\lortild',\lor^{-1}p)
\end{align} 
for all $\lortild,\lortild'\in\Lortild$ and $p\in\Hyp.$ 
\paragraph{Proper Poincar\'e Group.} 
The proper Poincar\'e group $\Poj$ can be obtained from the proper
orthochronous Poincar\'e group by adjoining the reflection $j$ at the
$x^2$-axis, {\it cf.\ }equation~\eqref{eqJ}, with the appropriate relations:   
\begin{align} 
 j^2&=\unit   &  j\,(a,\unit)\,j&= (j\act a,\unit)\,\nonumber\\
 j\,\Boox{t}\,j&= \Boox{t}  &  j\,\Rot{\omega}\,j&=\Rot{-\omega}. 
\label{eqjlorj}
\end{align}
(Note that the last equations imply $j\Boo{2}{t}j= \Boo{2}{-t}.$)  
Correspondingly, the universal covering group $\Potildj$ 
of this (disconnected)  group may be defined by adjoining an element 
$\jtild$ to $\Potild$ satisfying the relations 
\begin{equation}\label{eqjtild}
\jtild^2=\unit\quad\mbox{ and }\quad
\jtild\,\left(a,(\gamma,\omega)\right)\,\jtild =
 \left(j a,(\bar\gamma,-\omega)\right)\,. 
\end{equation}
In fact, the map $\jtild\mapsto j,$
$\lortild\mapsto\lor$ is a homomorphism and hence a covering projection. 
Finally, we prove an important cocycle relation of the Wigner rotation
\eqref{4.8b} with respect to $\jtild$. 
\begin{Lem} \label{WigJ}
{}For all $\lortild\in\Lortild$ and $p\in\Hyp$ the following relation
holds: 
\begin{equation}  \label{eqWigj}
 \Omega\big(\jtild\lortild\jtild\,,\,p\big)=-\Omega(\lortild\,,-j\act p)\;. 
\end{equation}
\end{Lem}
\begin{Proof}
From the definition of $\tilde h(p)$ via equation~\eqref{eqgamp} and the
group relations~\eqref{eqjtild} satisfied by $\jtild$ we get 
\begin{equation} \label{eqh(-jp)}
 \tilde h(-j\act p)=\jtild\,\tilde h(p)\,\jtild\;. 
\end{equation}
This implies
$t(\lortild,-j\act p)=\jtild\,t(\jtild\lortild\jtild,p)\,\jtild$ and hence
the claim. 
\end{Proof}
\begin{Lem} \label{JUni} 
i)  Let $U$ be the irreducible representation of $\Potild$ for mass 
$m>0$ and spin $s\in\Bb$ defined in equation~\eqref{eqUms}, and let 
$J$ be the operator defined in equation~\eqref{eqUj}. Then $J$ is an 
anti-unitary involution satisfying the representation property 
\begin{equation} \label{eqUjUgUj}
 J \U(\potild)J =\U(\jtild\potild\jtild)\;.
\end{equation} 
ii) Let $U$ be a finite direct sum of copies of the irreducible 
representation of $\Potild$ for mass $m>0$ and spin $s\in\Bb$, 
acting on a Hilbert space $\calH$. 
Then there is a unique, up to equivalence, extension of $U$ from 
$\Potild$ to $\Potildj$ in $\calH$. Uniqueness means that if $J$ and
$\hat{J}$ are anti-unitary involutions satisfying the representation
property~\eqref{eqUjUgUj}, then there is a unitary $V$ commuting with
$U(\Potild)$ and satisfying $VJ=\hat{J}V$. 
\end{Lem}
\begin{Proof}
i) follows immediately from Lemma~\ref{WigJ}.  
ii) The existence of $J$ follows from i) by taking direct
sums. To see uniqueness, let $\CC\doteq \hat{J} J$. It is a unitary 
operator commuting with the
representation $U$ and satisfying $\CC J=J\CC^{-1}$. Using 
spectral calculus in the same way as in the proof of Prop.\ 3.1 
in~\cite{Wol00}, we define a unitary root 
\newcommand{\Chalf}{V} 
$\Chalf$ of $\CC$, $\Chalf^2=\CC, $ which 
still commutes with the representation $U$ and satisfies $ \Chalf 
J=J \Chalf^{-1}$. Then $V$ has the properties claimed in the Lemma. 
\end{Proof}
\paragraph{Action of $\Potildj$ on $\Cccpaths$.} 
The universal covering group $\Potildj$ of the proper Poin\-car\'e
group  acts on $\Cccpaths$ in the following way. 
Let $\cccpath=(\ccc,\spdpath)\in\Cccpaths$ where $\spdpath$ is the equivalence 
class w.r.t.\ $\ccc$ of a path $t\mapsto \spdpath(t)$ in $\Spd$ starting at
$\refspd$ and ending in $\ccc$. 
Identifying $\Lortild$ with the set of homotopy classes of paths in
$\Lor$ starting at the unit, an element $\potild=(a,\lortild)\in\Potild$ 
acts on $\cccpath$ as follows. 
Let $t\mapsto\lortild(t)$ be any path in $\Lor$ which
represents $\lortild$. Then we define 
\begin{equation} \label{eqPoincSpc} 
 \potild\act\cccpath  \doteq \big(\po\act \ccc, 
\lortild\act \spdpath\big)\,,
\end{equation}
where $\lortild\act\spdpath$ is the equivalence class w.r.t.\ 
$\lor\act\ccc$ of the path $t\mapsto\lortild(t) \act \spdpath(t)$ in $\Spd$. 
Further, the element $\jtild\in\Potildj$ acts on $\cccpath$ as 
\begin{equation} \label{eqJSpc} 
 \jtild\act\cccpath \doteq \big(j \act \ccc, 
 \jtild \act \spdpath\big)\,,
\end{equation}
where $\jtild \act \spdpath$ is the equivalence class w.r.t.\ 
$j\act\ccc$ of the path $t\mapsto j \cdot \spdpath(t)$. Note
that this path also starts at $\refspd$ since we have chosen the
reference direction $\refspd$ so as to be invariant under $j$. 
\section{Proof of Analyticity of the Cocycle.}  \label{proof}
We establish the required analyticity properties of the cocycle 
$c(\lortild,p)$, {\it cf.\ }equ.~\eqref{DefCoc}, starting with 
the Wigner rotation factor for the $1$-boosts. Let  
\begin{align}  
l(p)&\doteq p_0-p_1+m-ip_2\quad \text{ and } \label{eqLin} \\
v(p)&\doteq {l(p)}\,{\overline{l(p)}}^{-1} . \label{eqV}
\end{align}
Note that for all $p\in\Hyp,$ the number $v(p)$ lies in the cut complex plane 
$\Bc\setminus \Bb^-_0,$ allowing for our definition of the power 
$v(p)^s$ given before Lemma~\ref{CocW1}. We have 
\begin{Lem} \label{Oml1} 
The Wigner rotation factor for the $1$-boosts is given by 
\begin{align}  
e^{is\Omega(\boox{t},p)}&=
v(p)^s\,v(\boox{-t}p)^{-s}\;.\label{eqOmv}
\end{align}
As a function of $t$, it has branch points in the strip 
$\Bb+i(0,\pi)$ for any $p\in\Hyp$  if $s$ is not an integer. 
\end{Lem}
\begin{Proof}
Equation~\eqref{eqOmv} is verified by direct calculation. But using 
\begin{align} 
 (\Boox{t}p)_0-(\Boox{t}p)_1 =e^{t}(p_0-p_1)\,,\label{eqp0p1}
\intertext{we get}
v(\Boox{-t}\,p) = \frac{e^{t}(p_0-p_1)+m-ip_2}{e^{t}(p_0-p_1)+m+ip_2}\,. 
   \label{eqVLam} 
\end{align} 
{}For any fixed $p\in\Hyp$, this function has zeroes in the strip, which
proves the claim. 
\end{Proof}
 
In the next proposition, we give an explicit expression for the  
cocycle $c(\lortild,p)$, exhibiting its analyticity properties. 
\begin{Prop} \label{Coc}
Let $\lortild=\boox{t}\;\boo{2}{t'}\;\rot{\omega}$, with 
$t,t',\omega\in\Bb$,  and let 
$\omega'\doteq\omega-\frac{\pi}{2}.$ Let further $p\in\Hyp$ be
arbitrary. 
\\ 
i) Denote by $\omega_0'$ the representant of $\omega'+2\pi\Bi$ in 
the interval $(-2\pi,0].$ Then 
\begin{align} \label{eqCoc}
\coc(\lortild,p) &=
2^{-s}\,e^{s(t+t')}\, e^{is(\omega'-\omega_0')}\,
\Big(a-b + e^{-t}(a+b)\,
\frac{-p_2+im}{p_0-p_1}\Big)^{2s}\;, \quad\text{ where }\\
a&\doteq \cos\frac{\omega_0'}{2}\;,
\quad \text{ and } \quad b\doteq e^{-t'}\sin\frac{\omega_0'}{2}\leq 0\,.   
 \label{eqab}  
\end{align} 
The power of $2s$ is understood within $\Bc\setminus \Bb_0^-$.  
% \pagebreak 
 \\
ii) Let $s\not\in\half\Bn_0$. The function 
$ \tau\mapsto \coc(\boox{\tau}\lortild,p)$  
is analytic in the strip $\Bb+i(0,\pi)$ if and only
if the parameters $t'$ and $\omega'$ satisfy the relation 
\begin{equation} \label{eqCocAna}
-\sinh{t'}\,|\sin{\omega'}|\leq \cos{\omega'}\,.
\end{equation}
In this case, the upper and lower boundary values are related by 
\begin{equation} \label{eqCocipi}
  \coc(\boox{\tau}\lortild,p)|_{\tau=i\pi}=e^{i\pi
    s}\,e^{4\pi isn(\omega')} \,\overline{\coc(\lortild,-jp)}\;,
\end{equation}
where $n(\omega')$ is the unique integer such that $\omega'-2\pi 
n(\omega')\in(-\pi,\pi].$ 
 \\
iii)  For $s\in\half\Bn_0$, the function 
$ \tau\mapsto \coc(\boox{\tau}\lortild,p)$  
is analytic in the strip $\Bb+i(0,\pi)$  and satisfies the boundary 
condition~\eqref{eqCocipi} for {\em all} $\lortild\in\Lortild$ . 
\end{Prop} 
\begin{Remark} From $(iii)$ follows that for $s\in\half\Bn_0$ the
  localization structure can be non-trivially extended to bounded
  regions as in Proposition~\ref{SFT}.  
The same  can be shown for $s\in-\half\Bn$ if one uses, instead of our
  intertwining function $u =: u_{s}$ the function $u^-_s(p) := 
  \overline{u_{|s|}(p)}$.  
\end{Remark}
\begin{Proof}
In the following, $p$ denotes an arbitrary point on the mass shell. 
We will use the cocycle identity 
\begin{equation} \label{eqCocRel}
 \coc(\lortild\lortild',p)=\coc(\lortild,p)\;\coc(\lortild',\lor^{-1}p) \,,
\quad \lortild,\lortild'\in\Lortild\,,
\end{equation}
satisfied by $\coc$ 
as a consequence of equation~\eqref{eqOmCoc}. 
Thus, we  first calculate $\coc(\lortild,p)$ if $\lortild$ is a boost
in $1$-direction or a rotation, 
and then use the above cocycle property  for a general element $\lortild$. 

The function $v$ from Lemma~\ref{Oml1} is related to $u,$ defined 
in equation~\eqref{equ}, by 
\begin{equation*}
  u(p) =\big(\frac{p_0-p_1}{m}\big)^{s} \,v(p)^s. 
\end{equation*}
Hence, in view of the identity~\eqref{eqp0p1}, Lemma~\ref{Oml1} implies that 
\begin{equation} \label{eqCocl1}
 \coc(\boox{t},p) = e^{st} \quad \text{ for all }t\in\Bb, p\in\Hyp \;.
\end{equation}
In order to calculate the cocycle for rotations, let us see how the
function $u$ transforms under rotations. Writing $u$ as 
\[ 
u(p) = \big(\frac{p_0-p_1}{m}\big)^s\,
\big({\lin(p)}/{\overline{\lin(p)}}\big)^{s}\,
\]
and using the identity 
\begin{equation*} 
\lin(p)\cdot\overline{\lin(p)}\,=\, 2(p_0+m)(p_0-p_1)\;,
\end{equation*} 
we get 
\[u(p)=\big(2m(p_0+m)\big)^{-s}\;\lin(p)^{2s}\;.\] 
Here we have used the fact that ${\rm Re }\,l(p)>0$ to identify
$(\lin(p)^2)^s$ with $\lin(p)^{2s}$.
A straightforward calculation shows that $\lin$ transforms under
rotations as follows: For $ \omega\in\Bb$, 
\begin{align} \label{eqllom}
\lin\big(\Rot{-\omega}p\big)&=\lin(p)\cdot\lin_{\omega}(p) \quad
 \text{ where } \\
\lin_\omega(p)&\doteq e^{-i\omega/2}\big(\cos\frac{\omega}{2}+ 
\sin\frac{\omega}{2}\,\frac{-p_2+im}{p_0-p_1}\big)\;.  \label{eqlom}
\end{align}
Note that $\lin(p)$ and $\lin_\omega(p)$ are, as well as the l.h.s.\ 
of equation~\eqref{eqllom},  in $\Bc\setminus \Bb_0^-$ for all $\omega$ 
and $p$. Hence we may take them to the power of $2s$ 
(within $\Bc\setminus \Bb_0^-$) separately, i.e.\
$(\lin(p)\,\lin_{\omega}(p))^{2s}=\lin(p)^{2s}\lin_{\omega}(p)^{2s}.$ 
We thus have  
\begin{align} \label{equlom}
u\big(\Rot{-\omega}p\big)&=u(p)\cdot\lin_{\omega}(p)^{2s} \,,
\intertext{and hence the cocycle for rotations is given by } 
  \coc(\rot{\omega},p) &= 
 e^{is\omega}\,\lin_\omega(p)^{2s}\,. \label{eqCocr} 
\end{align}
Now our results~\eqref{eqCocl1} and \eqref{eqCocr} imply, by the
cocycle relation~\eqref{eqCocRel}, that 
for all $t,\omega\in\Bb$  
\begin{equation} \label{eqCocl1r'}
  \coc(\boox{t}\rot{\omega},p) = e^{st}\,e^{is\omega}
  \lin_\omega(\Boox{-t}p)^{2s}\,. 
\end{equation}
Let us discuss how to take $\lin_\omega(p)$, see
equation~\eqref{eqlom}, to the power of $2s$. 
As is clear from the construction, the dependence of $\lin_\omega$ on 
$\omega$ is $2\pi$-periodic. Choosing a representant $\omega_0$ of 
$\omega+2\pi\Bi$ in the interval 
$(-2\pi,0]$, we may extract a factor $e^{-i s \omega_0}$ from 
$\lin_{\omega_0}(p)^{2s}.$  That is to say, we have 
\begin{equation}
 \lin_{\omega}(p)^{2s} = \lin_{\omega_0}(p)^{2s}=
  e^{-is\omega_0} \big(\cos\frac{\omega_0}{2}+ 
\sin\frac{\omega_0}{2}\,\frac{-p_2+im}{p_0-p_1}\big)^{2s}\,,\;\; 
\omega_0\in(-2\pi,0]\,.   
\end{equation} 
(For $\omega_0\neq 0$ this is so because then
the imaginary parts of the two factors on the r.h.s.\ of
equation~\eqref{eqlom} have opposite sign, while for
$\omega_0=0$ both factors equal one.) 
Using this and equation~\eqref{eqp0p1}, we arrive at  the expression 
\begin{equation} \label{eqCocl1r}
  \coc(\boox{t}\rot{\omega},p) = e^{st}\,e^{is(\omega-\omega_0)} \Big\{
\cos\frac{\omega_0}{2}+ e^{-t} 
\sin\frac{\omega_0}{2}\,\frac{-p_2+im}{p_0-p_1}\Big\}^{2s}\,,
\end{equation}
where $\omega_0$ is the representant of $\omega+2\pi\Bi$ in the
interval $(-2\pi,0].$

We are now prepared to prove equation~\eqref{eqCoc}. Let
$\lortild\in\Lortild$ be as in the Proposition. Using 
$\boo{2}{t'}=\rot{\pihalf}\,\boox{t'}\,\rot{-\pihalf},$ we rewrite
$\lortild$ as 
\begin{align} \label{eql1rl1r}
 \lortild = 
\boox{t}\,\rot{\pihalf}\,\boox{t'}\,\rot{\omega'}\;,\quad\text{ with
  }\;\omega'\doteq \omega-\pihalf\,. 
\end{align}
Due to the cocycle relation~\eqref{eqCocRel}, $\coc(\lortild,p)$ consists
of two factors of the form calculated in equation~\eqref{eqCocl1r}:  
\begin{align}  \nonumber 
\coc(\lortild,p)&=\qquad \coc\big(\boox{t}\,\rot{\pihalf}\,,\,p\big)
\qquad\,\cdot\,\qquad
\coc\big(\boox{t'}\,\rot{\omega'}\,,\,\Rot{-\pihalf}\Boox{-t}\,\,p\big) \\
&=2^{-s}\,e^{st}\,\Big\{1+e^{-t} \frac{-p_2+im}{p_0-p_1}\Big\}^{2s}\cdot 
e^{st'}e^{is(\omega'-\omega_0')}\,\Big\{a+b \;  
\frac{-q_2+im}{q_0-q_1}\Big\}^{2s}\,,
 \label{eqCocCoc} 
\end{align}
where $\omega_0'$ is the representant of $\omega'+2\pi\Bi$ in 
$(-2\pi,0],$  and we have written $a$ and $b$ as in equation~\eqref{eqab} 
of the Proposition and $q\doteq\Rot{-\pihalf}\Boox{-t}\,p$. 
Explicitely, $q$ reads 
\begin{align*}
q =
\big(\cosh t\,p_0-\sinh t\,p_1\,,\,p_2\,,\,\sinh t \,p_0-\cosh t \,p_1 \big), 
\intertext{and we calculate  }
\frac{-q_2+im}{q_0-q_1}=\frac{-e^tp_-+e^{-t}p_++2im}{e^tp_-+e^{-t}p_+ -2p_2} 
= -\frac{e^tp_-+ p_2-im}{e^tp_--p_2+im}\;,
\end{align*}
where $p_\pm\doteq p_0\pm p_1$.  
Then the product 
of the two curly brackets in \eqref{eqCocCoc} yields 
\begin{align} \label{eqc1c2}
\Big\{1+e^{-t}\frac{-p_2+im}{p_0-p_1}\Big\}\, 
\Big\{a+b \;  \frac{-q_2+im}{q_0-q_1}\Big\}
= a-b+e^{-t}(a+b)\, \frac{-p_2+im}{p_0-p_1}\,.
\end{align}
Having chosen $\omega_0'\in(-2\pi,0]$, we observe that $b\leq0,$ and
equality holds only if  $\omega_0'=0.$  Hence $a+b=0$ implies $a-b=1.$
Thus the r.h.s.\ is in $\Bc\setminus \Bb_0^-$. 
The same holds for the two factors on the l.h.s., hence we may take
them to the power of $2s$  (within $\Bc\setminus \Bb_0^-$) separately. 
We therefore have 
\begin{align} \label{eqCoc'} 
\coc(\lortild,p) &= 
2^{-s}\,e^{s(t+t')}\, e^{is(\omega'-\omega_0')}\,
f(t,p)^{2s}\,, \quad \text{ where }  \\
f(t,p)&\doteq a-b +
e^{-t}\,(a+b)\,\frac{-p_2+im}{p_0-p_1} \,.\label{eqCocz}
\end{align}
This proves part $i)$ of the Proposition. 
 
We now discuss the analyticity properties of the function 
$\coc(\boox{\cdot}\lortild,p).$  
If $\lortild$ is para\-met\-ri\-zed by $t,t',\omega\in\Bb$ as in the
Proposition, then 
$\boox{\tau}\lortild$ $=$ $\boox{\tau+t}\,\boo{2}{t'}\,\rot{\omega}$ and 
we may write 
\begin{align} \label{eqCocCoc'} 
\coc(\boox{\tau}\lortild,p) &=
2^{-s}\,e^{s(\tau+t+t')}\, e^{is(\omega'-\omega_0')}\,
f(\tau+t,p)^{2s}\;,
\end{align}
with $f(\cdot,p)$ as in equation~\eqref{eqCocz}. Note that
$f(\cdot,p)$ is an entire analytic function and satisfies 
\begin{equation} \label{eqZipi}
 f(t+i\pi,p) = \overline{f(t,-jp)}\;.
\end{equation}
{}For $s\in\half\Bn_0$ ($iii$), the claimed analyticity and boundary conditions
follow. To prove $ii)$, let $s\not\in\half\Bn_0$. Then  the 
function $\tau\mapsto \coc(\boox{\tau}\lortild,p)$ has an
analytic extension into the strip $\Bc+i(0,\pi)$ if and only if 
$f(\cdot,p)$ has no zeroes  in the strip. This can be decided by looking at 
the definition~\eqref{eqCocz}, taking into consideration
that $({-p_2+im})({p_0-p_1})^{-1}$ takes all values in the upper half
plane $\Bb+i\Bb^+$ if $p$ runs through $\Hyp.$ 

In the following,  $z_+^{2s}$ will denote 
$z$ to the power of $2s$ defined via the branch of 
the logarithm on $\Bc\setminus \Bb_0^+$  satisfying $\log(-1)=i\pi$, if 
$z\in\Bc\setminus\Bb_0^+$. For $z\in \Bc\setminus \Bb_0^-$, $z$ to the 
power of $2s$ defined via the branch of 
the logarithm on $\Bc\setminus \Bb_0^-$  satisfying $\log(1)=0$ will
now  be denoted by $z_-^{2s}$, instead of $z^{2s}$ as before.  
We will use the following rules:  
{\bf (1)} If $z$ is in the upper complex half plane, then $z_-^{2s}=z_+^{2s},$
while for $z$ in the lower half plane, $z_-^{2s}=e^{-4\pi is}z_+^{2s}.$ 
{\bf (2)} Complex conjugation  commutes with taking powers within 
$z\in \Bc\setminus \Bb_0^-$: $(\bar{z})_-^{2s}=\overline{z_-^{2s}}.$   
{\bf (3)} If $f(\tau,p)$ is contained in $\Bc\setminus \Bb_0^\pm$ for all 
$\tau$ in the strip $\Bb+i[0,\pi]$, then analytic continuation in
$\tau$ commutes with taking powers within $\Bc\setminus
\Bb_0^\pm$, respectively. That means in particular, 
$f(\tau,p)_{\pm}^{2s}|_{\tau=i\pi}=f(i\pi,p)_\pm^{2s}$, where the
l.h.s.\ denotes the analytic continuation of 
$ f(\cdot,p)_{\pm}^{2s}$ from the real line to $i\pi.$ 

{\bf Case 1:} $|b|>|a|.$ Then $(a+b)(a-b)<0,$ hence $a+b$ and $a-b$ have
different sign. Then $f(\cdot,p)$ has zeroes in the
strip and hence the cocycle has, for $s\not\in \half\Bi,$ no analytic 
continuation into the strip. 
{\bf Case 2:} $|b|\leq |a|,$ i.e.\ $(a+b)(a-b)\geq0.$ We observe first that $a=0$ 
implies $\omega_0'=-\pi,$ hence $b=-e^{-t'}<0,$ contradicting the
assumption. Hence $a\neq0$ in the present case. 
{\bf Case 2.1:} Both $a+b$ and $a-b$ are greater or equal to zero. 
Since $a\neq0$ (as observed above), this implies that $a>0$ and
consequently, $b$ being non-positive (cf.~\eqref{eqab}) that $a-b>0$. Hence 
$f(\tau,p)$ is contained in $\Bc\setminus \Bb_0^-$ for all $\tau$ in the
strip, and 
our rules above, together with equation~\eqref{eqZipi}, imply that 
$f(\tau,p)_-^{2s}|_{\tau=i\pi}=\overline{f(0,-jp)_-^{2s}}$. 
Hence we have 
\begin{equation} \label{eqCocFall2}
 \coc(\boox{\tau}\lortild,p)|_{\tau=i\pi}=e^{i\pi s}
\,e^{2is(\omega'-\omega_0')}\, \overline{\coc(\lortild,-jp)}\;.   
\end{equation}
In the case at hand, $a>0$ and consequently $\omega_0'\in(-\pi,0]$. 
Hence $(\omega'-\omega_0')/2\pi$ is just the 
integer $n(\omega')$ defined in the Proposition, and the above 
equation coincides with  equation~\eqref{eqCocipi}.  
{\bf Case 2.2:}  Both $a+b$ and $a-b$ are less or equal to zero. 
Similarly as in case 2.1, this implies that $a+b<0$. Hence 
$f(\tau,p)$ is in the lower half 
plane for real $\tau$, and is contained in 
$\Bc\setminus \Bb_0^+$ for all $\tau$ in the strip. Hence  our three
rules above imply that 
$f(\tau,p)_-^{2s}|_{\tau=i\pi}=e^{-4\pi is}\overline{f(0,-jp)_-^{2s}}$. 
We thus have 
\begin{equation} \label{eqCocFall1}
 \coc(\boox{\tau}\lortild,p)|_{\tau=i\pi}=e^{i\pi s}
\,e^{2is(\omega'-\omega_0'-2\pi)}\, \overline{\coc(\lortild,-jp)}\;.   
\end{equation}
In the case at hand, $a<0$ and consequently $\omega_0'\in(-2\pi,-\pi)$. 
Hence $(\omega'-\omega_0'-2\pi)/2\pi$ is just the integer $n(\omega')$
defined in the Proposition, and the above equation again coincides with 
equation~\eqref{eqCocipi}.  

We have now shown that the cocycle has an analytic continuation into
the strip  if and only if $|b|\leq|a|$, and that the continuation satisfies 
equation~\eqref{eqCocipi}. It remains to show that $|b|\leq|a|$ is equivalent 
to the condition~\eqref{eqCocAna}. 
Both conditions are true for $\omega'\in2\pi\Bi$ and false for 
$\omega'\in \pi+2\pi\Bi,$ hence they coincide if $\omega'\in \pi\Bi.$ 
If $\omega'\not\in\pi\Bi,$ then $|b|\leq|a|$ is equivalent to
$$e^{-t'}-e^{t'}\leq |\cot{\frac{\omega_0'}{2}}|-|\tan{\frac{\omega_0'}{2}}|
=2\cos\omega_0'\,|\sin\omega_0'|^{-1}=2\cos\omega'\,|\sin\omega'|^{-1}\,,$$  
hence to condition~\eqref{eqCocAna}. 
We have thus shown part $ii)$ of the Proposition. 
\end{Proof}
\subsection*{Acknowledgements} 
I thank B.~Schroer, H.-W.~Wiesbrock, D.~Guido and R.~Longo for stimulating 
discussions and D.~Buchholz for valuable hints concerning the manuscript.  
Further, I gratefully acknowledge the hospitality extended to me by 
the Universities of Rome~I and II, 
and financial support by the SFB 288 
(Berlin), the EU (via TMR networks in Rome), the Graduiertenkolleg 
``Theoretische  Elementarteilchenphysik'' (Hamburg), and the DFG
(G\"ottingen).  

\providecommand{\bysame}{\leavevmode\hbox to3em{\hrulefill}\thinspace}
\providecommand{\MR}{\relax\ifhmode\unskip\space\fi MR }
\providecommand{\MRhref}[2]{%
  \href{http://www.ams.org/mathscinet-getitem?mr=#1}{#2}
}
\providecommand{\href}[2]{#2}


\begin{thebibliography}{10}

\bibitem{Ban}
R.~Banerjee, A.~Chatterjee, and V.~V. Sreedhar, \emph{Canonical quantization
  and gauge invariant anyon operators in {Chern-Simons} scalar
  electrodynamics}, Ann. Phys. \textbf{222} (1993), 254--290.

\bibitem{Ba1}
V.~Bargmann, \emph{Irreducible unitary representations of the {Lorentz} group},
  Ann. Math. \textbf{48} (1947), 568--640.

\bibitem{BiWi}
J.J. Bisognano and E.H. Wichmann, \emph{On the duality condition for a
  {Hermitean} scalar field}, J. Math. Phys. \textbf{16} (1975), 985.

\bibitem{BGL}
R.~Brunetti, D.~Guido, and R.~Longo, \emph{Modular localization and {W}igner
  particles}, Rev.\ Math.\ Phs. \textbf{14} (2002), 759--786,
  arXiv:math-ph/0203021.

\bibitem{Bu87}
D.~Buchholz, \emph{On particles, infraparticles and the problem of asymptotic
  completeness}, Proc. IAMP Conf. (Marseille), World Scientific, 1987.

\bibitem{BuEp}
D.~Buchholz and H.~Epstein, \emph{Spin and statistics of quantum topological
  charges}, Fysica \textbf{17} (1985), 329--343.

\bibitem{BuF}
D.~Buchholz and K.~Fredenhagen, \emph{Locality and the structure of particle
  states}, Commun. Math. Phys \textbf{84} (1982), 1--54.

\bibitem{Ply}
J.L. Cortes and M.S. Plyushchay, \emph{Anyons as spinning particles}, Int. J.
  Mod. Phys. \textbf{A11} (1996), 1427.

\bibitem{FasS02}
L.~Fassarella and B.~Schroer, \emph{Wigner particle theory and local quantum
  physics}, J. Phys. A \textbf{35} (2002), 9123--9164, arXiv:hep-th/0112168.

\bibitem{F89}
K.~Fredenhagen, \emph{Structure of superselection sectors in low dimensional
  quantum field theory}, Proceedings (Lake Tahoe City) (L.L. Chau and W.~Nahm,
  eds.), 1989.

\bibitem{FM1}
J.~Fr\"{o}hlich and P.A. Marchetti, \emph{Quantum field theories of vortices
  and anyons}, Commun. Math. Phys. \textbf{121} (1989), 177--223.

\bibitem{Grig}
D.R. Grigore, \emph{Free fields for any spin in 1+2 dimensions}, Journ. Math.
  Phys. \textbf{35} (1994), 6304--6331.

\bibitem{H96}
R.~Haag, \emph{Local quantum physics}, second ed., Texts and Monographs in
  Physics, Springer, Berlin, Heidelberg, 1996.

\bibitem{Ito01}
H.~Ito, \emph{Statistics of the composite systems and anyons in the fractional
  quantum {H}all effect}, Prog. Theor. Phys. \textbf{107} (2002), 177--189.

\bibitem{Jack}
R.~Jackiw and E.J. Weinberg, \emph{Selfdual {C}hern-{S}imons vortices}, Phys.
  Rev. Lett. \textbf{64} (1990), 2234.

\bibitem{LRT}
P.~Leyland, J.~Roberts, and D.~Testard, \emph{Duality for quantum free fields},
  unpublished notes, 1978.

\bibitem{Mintchev91}
M.~Mintchev and M.~Rossi, \emph{Gauss law and charged fields in the presence of
  a {C}hern- {S}imons term}, Phys. Lett. \textbf{B271} (1991), 187--195.

\bibitem{M}
J.~Mund, \emph{No-go theorem for `free' relativistic anyons in $d=2+1$}, Lett.
  Math. Phys. \textbf{43} (1998), 319--328.

\bibitem{M01a}
\bysame, \emph{The {B}isognano-{W}ichmann theorem for massive theories}, Ann.
  H. Poinc. \textbf{2} (2001), 907--926.

\bibitem{Pablo}
P.~Ramacher, \emph{Modular localization of elementary systems in the theory of
  {W}igner}, J. Math. Phys. \textbf{41} (2000), no.~9, 6079--6089.

\bibitem{Re}
K.-H. Rehren, \emph{Field operators for anyons and plektons}, Commun. Math.
  Phys \textbf{145} (1992), 123.

\bibitem{RvD}
M.A. Rieffel and A.~Van~Daele, \emph{A bounded operator approach to
  {T}omita-{T}akesaki theory}, Pacific Journ. Math. \textbf{69} (1977), no.~1,
  187--221.

\bibitem{S63}
B.~Schroer, \emph{Infrateilchen in der {Q}uantenfeldtheorie}, Fortsch. Phys.
  \textbf{173} (1963), 1527.

\bibitem{S97a}
\bysame, \emph{{W}igner representation theory of the {P}oincar\'e group,
  localization, statistics and the {S}-matrix}, Nuclear Phys. \textbf{B 499}
  (1997), 519--546.

\bibitem{Semenoff}
G.W. Semenoff, \emph{Canonical quantum field theory with exotic statistics},
  Phys. Rev. Lett. \textbf{61} (1988), no.~5, 517.

\bibitem{Swanson}
M.S. Swanson, \emph{{Fock-Space} representations of coupled {Abelian}
  {Chern-Simons} theory}, Phys. Rev. \textbf{42} (1990), no.~2, 552.

\bibitem{Wig}
E.P. Wigner, \emph{On unitary representations of the inhomogeneous {Lorentz}
  group.}, Ann. Math. \textbf{40} (1939), 149.

\bibitem{Wol00}
M.~Wollenberg, \emph{Notes on modular conjugations of von {N}eumann factors},
  Z.\ Anal.\ Anwendungen \textbf{19} (2000), 13--22.

\end{thebibliography}
\end{document}